\documentclass[preprint,showpacs,preprintnumbers,amsmath,amssymb]{revtex4}

\usepackage{graphicx}
\begin{document}

\title{Coherent State Description of the Ground State in the Tavis-Cummings Model and its Quantum Phase Transitions}

\author{Octavio Casta\~nos$^{1}$, Ram\'on L\'opez-Pe\~na$^{1}$, Eduardo Nahmad-Achar$^{1}$, Jorge G. Hirsch$^{1}$, Enrique L\'opez-Moreno$^{2}$ and Javier E. Vitela$^{1}$}

\affiliation{$^{1}$Instituto de Ciencias Nucleares, Universidad Nacional Aut\'onoma de M\'exico,
Apdo. Postal 70-543 M\'exico 04510 D.F. \\ $^{2}$Facultad de Ciencias, Universidad Nacional Aut\'onoma de M\'exico, Apdo. Postal 70-542 M\'exico 04510 D.F.}

\begin{abstract}

Quantum phase transitions and observables of interest of the ground state in the Tavis-Cummings model are analyzed, for any number of atoms, by using a tensorial product of coherent states. It is found that this ``trial" state constitutes a very good approximation to the exact quantum solution, in that it globally reproduces the expectation values of the matter and field observables. These include the population and dipole moments of the two-level atoms and the squeezing parameter. Agreement in the field-matter entanglement and in the fidelity measures, of interest in quantum information theory, is also found.The analysis is carried out in all three regions defined by the separatrix which gives rise to the quantum phase transitions.  It is argued that this agreement is due to the gaussian structure of the probability distributions of the constant of motion and the number of photons. The expectation values of the ground state observables are given in analytic form, and the change of the ground state structure of the system when the separatrix is crossed is also studied.

\end{abstract}

\pacs{42.50.Ct, 64.70.Tg, 03.65.Fd}

\maketitle

\section{Introduction}

A model of fundamental importance in quantum optics and essential in any description of systems involving the interaction between matter and light is given by the Hamiltonian proposed by Dicke~\cite{dicke54}. This describes the interaction of N-identical two-level systems with a single mode radiation field in the dipole approximation; the particles are confined in a container small compared to the radiation wavelength.

The simplest completely soluble quantum-mechanical model of one atom in an electromagnetic field is described by the Jaynes-Cummings (JC) model~\cite{jc1963,dodonov,shore}. In this model the Hamiltonian terms which do not conserve the energy of the field plus the occupation number of the two-level system have been neglected because their contributions are very small except for very high intensity fields~\cite{jc1963}, which has been called the {\it rotating-wave approximation}.

Since the end of the sixties, the exact solution for an N-molecule -- radiation field Hamiltonian was obtained~\cite{tavis}, and a full family of exactly solvable generalizations has been presented recently~\cite{duke}. They have provided insight into other more complicated physical systems and offers a standard comparison for approximation techniques. Presently, the JC model and its generalization given by Tavis and Cummings, the {\it Tavis-Cummings} (TC) model, continue to be fundamental to study basic properties of quantum electrodynamics and to understand phenomena like the existence of collapse and revivals in the Rabi oscillations~\cite{eberly}, the formation of macroscopic quantum states, and the many experimental studies of Rydberg atoms with very large principal quantum number within single-mode cavities~\cite{dodonov,shore,haroche}. It is important to mention that it was not until 1987 that the collapse and revival predicted by the JC model was experimentally observed~\cite{rempe}.

Instabilities, or quantum phase transitions, in these models have recently been studied due to their influence on several kinds of entanglement, important in quantum information theory. Lambert et al.~\cite{lambert} have studied the atom-field entanglement in the thermodynamic limit, finding logarithmic divergences in the atom-field entanglement and discontinuities in the average linear entropy. Bu\v{z}ek et al.~\cite{buzek} showed that an infinite sequence of instabilities appear in the ground state of the Dicke model when the rotating-wave approximation is taken, characterized by changes in the bipartite entanglement between atoms. Similar models have been used to describe the phase transitions of $N$ identical qubits interacting with a bosonic mode in an adiabatic approximation~\cite{liberti}.

Although there have been recent contributions to the quantum phase transitions in the Dicke model and its extensions~\cite{reslen,zanardi,goto}, we want to enhance the role of the catastrophe formalism to determine when significant changes to the ground state occur for small changes of the external environment (the parameters of the model). According to Sachdev~\cite{sachdev}, if a system undergoes a ground state energy phase transition at $T=0$ as a function of external parameters, then it also has a thermodynamic phase transition for fixed interaction parameters as a function of the increasing temperature.

Quantum phase transitions and stability properties of algebraic models have been studied through the catastrophe formalism and the coherent states theory~\cite{gilmore3}. Recently, a procedure was established to determine the phase transitions associated to nuclear and matter models~\cite{cas01}. In this contribution we present a comprehensive study of the phase transitions of the ground state in the TC model, for any number of atoms, when the strength of the interaction $\gamma$ and the atom hyperfine two-level separation energy $\omega_A$ are varied. These quantum transitions occur also for a small number of particles, although we can take the thermodynamic limit and corroborate the results obtained at the beginning of the 70's~\cite{lieb}. The minima of the expectation value of the Hamiltonian are calculated for a trial state, built as a coherent state $HW(1) \otimes SU(2)$. Employing the catastrophe formalism, the separatrix is found. The structure of the ground state of the system changes when this separatrix, given by $\omega_A=\pm \gamma^2$, is crossed, in agreement with the phase transition found by Hepp and Lieb~\cite{lieb}.  These parabolae divide the control parameter space in three regions which map naturally onto the Bloch sphere, where minima of the expectation value of the Hamiltonian with respect to the coherent trial state are attained: (1) the North Pole, where the ground state of the system is constituted by all the atoms lying in the low hyperfine level and the electromagnetic field has zero photons; (2) the South Pole, with the ground state formed by all the atoms in the excited hyperfine level and zero photons in the electromagnetic field; and (3) the Parallel region, where the ground state has a variable number of components (field + matter), ranging from one, for values of the interaction strength close to the arms of the parabolae, to $N+1$, for the collective regime. By means of Ehrenfest's thermodynamic classification of phase transitions we have determined that for the TC model there are second order phase transitions when the separatrix is crossed through the arms of the parabolae, and first order transitions when the crossing happens through their vertices.
We also explore the influence of the phase transitions on the behaviour of observables of interest for the matter and the field: the population and dipole moments of the two level atoms, together with the expectation value of the number of photons and its fluctuations. We evaluate the squeezing and entanglement properties of the ground state in the different regions of the control parameter space. In order to compare the reduced probability distribution of the excited number of atoms, obtained by the exact quantum result, with that determined using the semi-classical approximation, the fidelity between them is calculated. Finally, the photon number distribution of the ground state as a function of the atom-field coupling parameter $\gamma$ is given.

\section{Semiclassical Analysis}

By {\it semiclassical analysis} it is meant here the following procedure: i) calculation of the expectation value of the Hamiltonian with respect to the tensorial product of coherent states ({\it trial state}); ii) determination of the minimal critical points; and iii) use of the catastrophe formalism to find the stability properties. The use of coherent states as trial states lets us determine in analytical form the expectation values of matter and field observables.

\subsection{Energy Surface and Critical Points}

\noindent
We are interested in solving the TC Hamiltonian
given by~\cite{jc1963}
	\begin{equation}
		H_{TC}=\omega_{F}\,a^{\dagger}a+\omega_{A}^{\prime}\,J_{z}
		+\frac{\gamma^{\prime}}{\sqrt{N}}\left(a^{\dagger}\,J_{-}
		+a\,J_{+}\right)\ ,
	\end{equation}
where $N$ is the number of particles in the system, $\omega'_F$ is the field frequency, $\omega'_A$ the atomic energy-level difference, and $\gamma'$ the field-atom interaction strength. It is immediate that this Hamiltonian commutes with the operator 
	\begin{equation}
		\Lambda = a^{\dagger}a+J_{z}\ .
		\label{constant-motion}
	\end{equation}
It is then convenient to rewrite it
by introducing a detunning parameter $\Delta$, and by dividing it by 
$\omega_{F}$ (which can be thought of as the natural unit of frequency) and by the total number of particles, having in this way an intensive Hamiltonian operator
	\begin{equation}
		H=\frac{1}{N}\,\Lambda -  \frac{\Delta}{N}\,J_{z}
		+\frac{\gamma}{\sqrt{N}\,N}\left(a^{\dagger}\,J_{-}
		+a\,J_{+}\right)\ ,
		\label{jaynes-cummings}
	\end{equation}
where $\Delta= 1 - \frac{\omega^\prime_A}{\omega_F} \equiv 1 - \omega_A$ and $\gamma= \frac{\gamma^\prime}{\omega_F}$.

In order to obtain an energy surface we use as a trial state the direct product
of coherent states in each subspace: Heisenberg-Weyl for the photon
part~\cite{dodonov} and $SU(2)$ or spin for the particle part~\cite{gilmore1972}, i.e., $\vert\alpha, \,\zeta\rangle=\vert\alpha\rangle \otimes\vert\zeta\rangle$. Explicitly, it is given by the expression
	\begin{equation}
		\vert\alpha ,\,\zeta\rangle =\frac{\exp\left(-\left|\alpha\right|^{2}/2\right)}{
		\left(1+\left|\zeta\right|^{2}\right)^{j}}
		\sum_{\nu=0}^{\infty}\sum_{m=-j}^{+j} \left\{\frac{\alpha^{\nu}}{\sqrt{\nu!}}\,
		\binom{2j}{j+m}^{1/2}\,
		\zeta^{j+m}\,\vert\nu\rangle \otimes \vert j,\,m\rangle \right\}\ ,
	\label{trial}
	\end{equation}
where the ket $\vert \nu \rangle$ is an eigenstate of the photon number operator, $\vert j,\, m \rangle$ is a Dicke state with $j$ denoting the eigenvalue of $J^2$, and $m$ the corresponding eigenvalue of $J_z$. The trial state contains $N$ particles and up to an infinite number of photons distributed into all the possible ways between the two levels. 

The expectation value of the TC Hamiltonian (\cite{tavis}) in this
trial state is defined as the energy surface given by
	\begin{eqnarray}
		E(q,\,p,\,\theta,\,\phi)&=&\langle\alpha, \,\zeta\vert\,H\,
		\vert\alpha, \,\zeta\rangle\nonumber\\
		&=&\frac{\lambda(j,q,\,p,\,\theta,\,\phi)}{N}+\Delta \frac{j}{N}\,\cos\theta
		+\frac{\gamma}{\sqrt{2 N}}\,\sin\theta\left(q\,\cos\phi-p\,\sin\phi
		\right)\ .
	\end{eqnarray}
To get this expression we have substituted
	\begin{eqnarray}
		\alpha&=&\frac{1}{\sqrt{2}}\left(q+i\,p\right)\ ,\\
		\zeta&=&\tan\left(\frac{\theta}{2}\right)\,\exp\left(i\,\phi\right)\ ,
		\label{dzeta}
	\end{eqnarray}
where $(p,q)$ correspond to the expectation values of the quadratures of the field and $(\theta,\phi)$ determine a point on the Bloch sphere. The function $\lambda$ is the expectation value of the constant of motion $\Lambda$, and is given by
	\begin{equation}
		\lambda(j,q,\,p,\,\theta,\,\phi)=\langle\alpha, \,\zeta\vert \Lambda 	       					    \vert\alpha,\,\zeta\rangle=\frac{1}{2}\left(q^{2}+p^{2}\right)
		- j\,\cos\theta\ .
	\end{equation}
The eigenvalue $j$ is related to the number of molecules or atoms present in the two-level system, and it can take the values $\vert m \vert \leq j \leq N/2$~\cite{gilmore1972}, according to the symmetry of the particles. If we are interested in the symmetric configuration, the so called superradiance regime~\cite{lieb}, we have to consider $N=2 j$, which will be assumed hereafter.

The critical points of the energy surface are obtained by equating its first derivatives to zero. The derivatives with
respect to $q$ and $p$ give us the relation of them with the critical values of $\theta$ and $\phi$:
	\begin{eqnarray}
		q_{c}&=&-\sqrt{j}\,\gamma\,\sin\theta_{c}\,\cos\phi_{c}\ ,
	\label{quadratureq}\\
		p_{c}&=&\phantom{-}\sqrt{j}\,\gamma\,\sin\theta_{c}\,\sin\phi_{c}\ .
	\label{quadraturep}
	\end{eqnarray}
The derivative with respect to $\theta$, using (9) and (10), determines its
critical value $\theta_c$ in terms of the strength parameter $\gamma$ and of $\omega_{A}$:
	\begin{equation}
		\sin\theta_{c}\, (\omega_{A} - \gamma^2 \,\cos\theta_{c}) =0\ .
	\end{equation}
The remaining equation, related with the derivative with respect to $\phi$, gives us an identity. One can check that the Hessian matrix has rank 3 at the critical points. This means that the energy surfaces at the critical points are independent of $\phi$ and we shall say the energy surface is $\phi$-unstable. 

We thus obtain the following critical points: the North Pole ($\theta_{c}=0$),
the South Pole ($\theta_{c}=\pi$), and a Parallel defined by
$\theta_{c}=\arccos\left(\omega_{a}/\gamma^{2}\right)$.

We determine the nature of the critical points examining the hessian
matrix of the energy surface at the critical points. Because the
variable $\phi$ is spurious, we have the liberty to choose its value (which in general we will take to be zero). We can recover the values of quantities dependent on $\phi$, at any $\phi$, by applying the transformation $e^{i\phi\Lambda}$. 
	
In the North Pole ($\theta_c = 0$) the hessian matrix has the three eigenvalues 
	\[ 
		1/(2\,j), \quad \left(1+j\,\omega_{A}\pm\sqrt{(1-j\,\omega_{A})^{2}
		+4\,j\,\gamma^2}\right)/(4\,j)\,.
	 \]
For $\omega_{A} > \gamma^2$ we have a minimum (all three eigenvalues positive); for $\omega_{A} = \gamma^2$ we obtain a degenerate critical point; and for $\omega_{A} < \gamma^2$ we get a saddle point (one eigenvalue is always negative). Along $\omega_{a}=\gamma^{2}$, the hessian matrix is singular and thus it is a bifurcation set of the system~\cite{gilmore3}.

In the South Pole ($\theta_c = \pi$) the hessian matrix has eigenvalues
	\[ 
		1/(2\,j), \quad \left(\pm(1-j\,\omega_{A})+\sqrt{(1+j\,\omega_{A})^{2}
		+4\,j\,\gamma^2}\right)/(4\,j)\,.
	\]
Here, for $\omega_{A} < -\gamma^2$ we have a minimum; for $\omega_{A} = -\gamma^2$ we obtain a degenerate critical point; and for $\omega_{A} > -\gamma^2$ we get a saddle point.  The matrix is singular along $\omega_{A} = -\gamma^2$, which is then another bifurcation set.

Finally, in the Parallel region ($\theta_c = \arccos(\omega_{A}/\gamma^{2})$),
the hessian matrix eigenvalues are
	\[
		1/(2\,j), \quad \left(\gamma^{2}(1+j\,\gamma^{2})\pm\sqrt{\gamma^{4}\,
		(1-j\,\gamma^{2})^{2}+4\,j\,\omega_{A}^{2}\,\gamma^{2}}\right)/(4\,j\,\gamma^{2})\,.
	\]
In this case,
there are no solutions above $\omega_{A}=\gamma^{2}$ nor below $\omega_{A}=-\gamma^{2}$, and in between all the eigenvalues are positive. Along $\omega_{a}=\pm\gamma^{2}$, the hessian matrix is singular.

\begin{figure}[h]
\scalebox{0.8}{\includegraphics{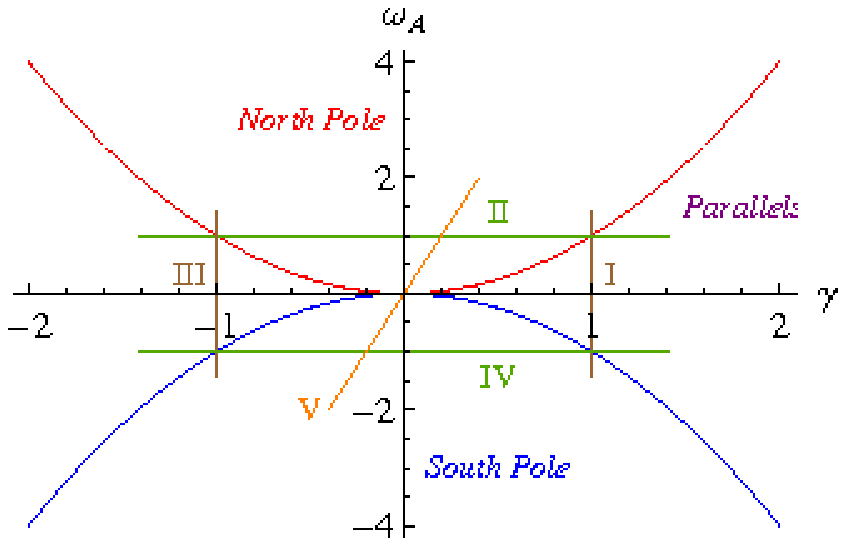}} \qquad
\scalebox{0.7}{\includegraphics{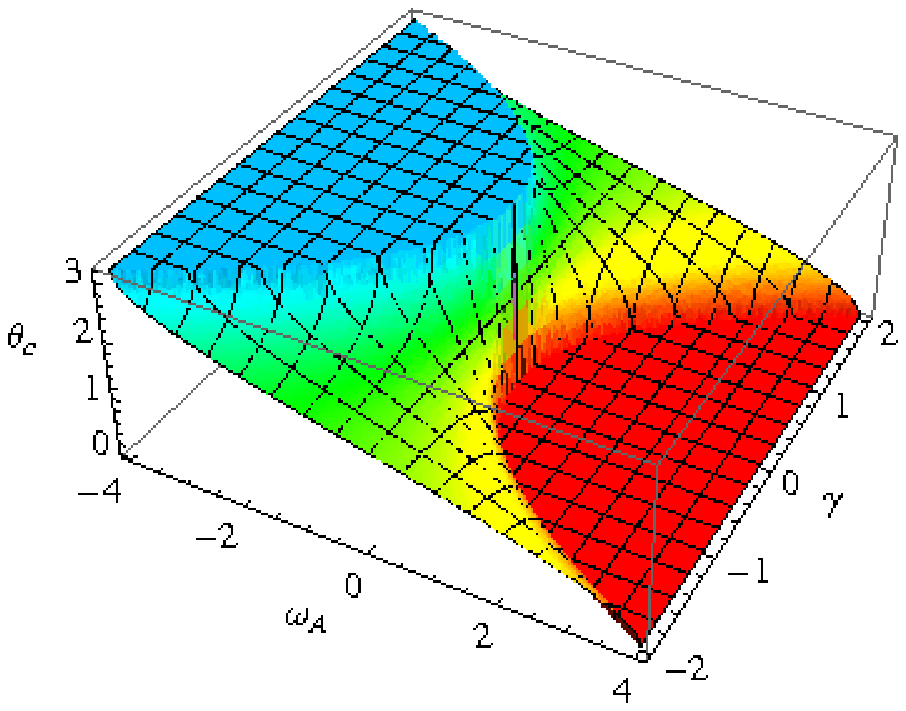}}
\caption{\label{separa} On the left, the separatrix of the system in the parameter space $(\gamma,\omega_A)$ is displayed. A trajectory in the control parameter space is shown, along which we analyze the behaviour of the semiclassical energy surface. On the right, $\theta_{c}$ at minima is shown as a function of the interaction strength $\gamma$ and $\omega_A$.}
\end{figure}

The minimum energy function therefore maps all the region $\omega_{A}>\gamma^{2}$ to the North Pole of the Bloch sphere, all the region $\omega_{A}<-\gamma^{2}$ to the South Pole, and the region $-\gamma^{2}<\omega_{A}<\gamma^{2}$ to the rest of the sphere. Along the curves $\omega_{A}=\pm\gamma^{2}$ the nature of the critical points is not determined by the hessian; they are bifurcation sets in the parameter space forming part of the separatrix of the system, which is shown in Fig.(\ref{separa}). 

Table \ref{critical} shows the critical points, the energy, and the constant of motion
evaluated at these, together with the conditions in the parameter space to guarantee that they constitute an energy minimum. 

\begin{table}
\caption{Expectation values of energy per number of particles and constant of motion, of the estimated ground state of the system. The last column shows the conditions for E to be a minimum in each region.}
\begin{center}
\begin{tabular}{|c||c|c|c|}
\hline
Minima&$\frac{E_0}{N}$& $ \lambda $ &Conditions\\
\hline\hline
$\theta_{c}=0$&\rule[-4mm]{0mm}{10mm}$-\frac{\omega_{A}}{2}$&$-j$ \quad& $\omega_{A}>\gamma^2$ \\
\hline
$\theta_{c}=\pi$&\rule[-4mm]{0mm}{10mm}$\phantom{-}\frac{\omega_{A}}{2}$\quad
&$\phantom{-}j$ & $\omega_{A}<-\gamma^2$\\
\hline
$\theta_{c}=\arccos\left(\omega_{a}/\gamma^{2}\right)$&
\rule[-4mm]{0mm}{10mm}$-\frac{\omega_{A}^{2}+\gamma^{4}}{4\,\gamma^{2}}$&\quad
$ j \, \frac{-\omega_{A}\,\left(\omega_{A}+2\right)+\gamma^{4}}{2\,\gamma^{2}}$ & $-\gamma^2<\omega_{A}<\gamma^2$\\
\hline
\end{tabular}
\end{center}
\label{critical}
\end{table}

\begin{figure}[h]
\scalebox{1.0}{\includegraphics{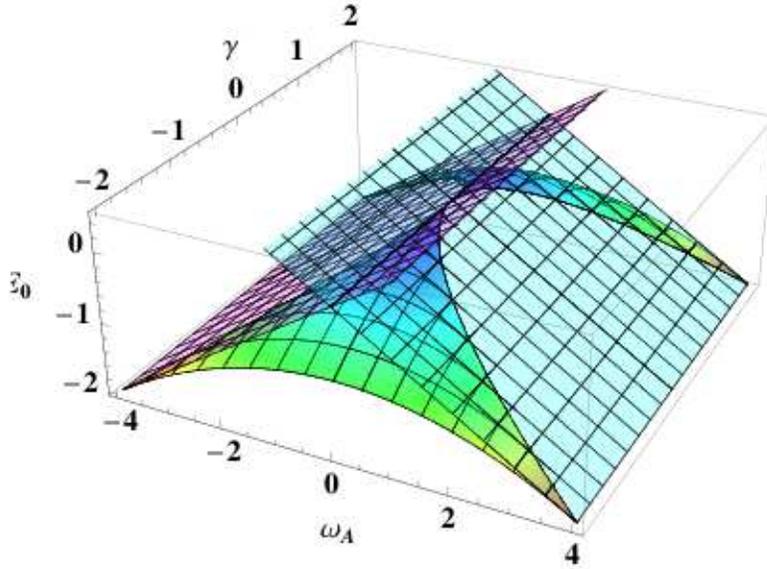}}
\caption{\label{EnergyPlot} Energy surface $E$ as a function of the interaction strength $\gamma$ and the atomic energy level difference $\omega_{A}$. The plane growing with $\omega_{A}$ represents $E$ at the North Pole, the other plane is $E$ at the South Pole, and the paraboloid is $E$ at the Parallels region. Note that the planes cut the paraboloid at $|{\omega_{A}}| = \gamma^{2}$; thus, the minimum energy is found at different regions depending on the relationship of $\omega_{A}$ to $\gamma$.}
\end{figure}

Most system variables will inherit their behaviour from that of $\theta_c$ at the minima. Fig.(\ref{separa}) shows $\theta_c$ as a function of the interaction strength $\gamma$ and $\omega_A$. The full energy surface is shown in Fig.(\ref{EnergyPlot}) as a function of $\gamma$ and $\omega_A$ also.

\subsection{Phase Transitions}

The order of the phase transitions can be determined following the
Ehrenfest classification. A phase transition takes place between $p$ and $q$ branches of critical
points, and is of nth-order if
	\begin{equation}
		\lim_{\delta \to 0} \left.
		\frac{\partial^{i} E_{0}^{(p)} (s)}{\partial s^{i}}
		\right|_{s_{0}-\delta} =
		\lim_{\delta \to 0} \left.
		\frac{\partial^{i} E_{0}^{(q)} (s)}{\partial s^{i}}
		\right|_{s_{0}+\delta} \ ,
	\end{equation}
for $i=0$, $1$, $2$, \dots , $n-1$, but fails for
$i=n$~\cite{gilmore3}. In the TC model, the  phase transitions for the states of mimimum energy occur at the
separatrix $\omega_{A}=\pm\gamma^{2}$, as we have seen. It can be shown that crossing the arms of the parabolae, such as along the circuit $I-II-III-IV-I$ in Fig.(\ref{separa}), leads to second order phase transitions; in this case the composition of the ground state changes from only one component $\vert 0 \rangle\otimes\vert j, \,-j \rangle$ at the North Pole to many components at the Parallels region, or from one component $\vert 0 \rangle\otimes\vert j,\, j \rangle$ at the South Pole to many at the Parallels, and viceversa. Figure (\ref{ContourII}) below shows contour maps for $E=E(\theta,\phi)$ just before and after the crossing at $\gamma=-1$ along Path II.

\begin{figure}[h]
\scalebox{1.0}{\includegraphics{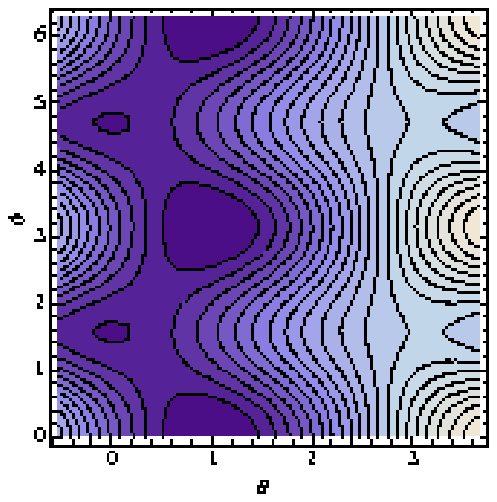}} \qquad
\scalebox{1.0}{\includegraphics{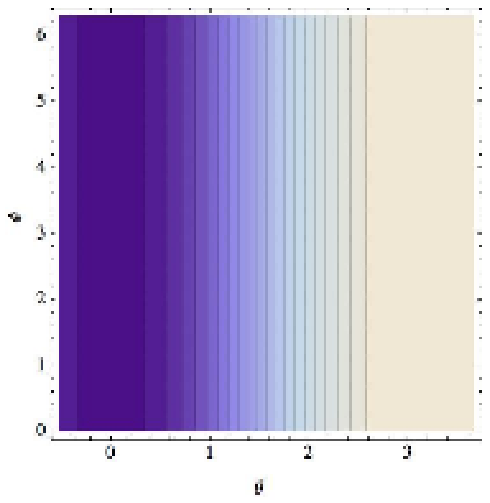}}
\caption{\label{ContourII} Contour levels of the surface energy function $E(\theta,\phi)$ near the arms of the separatrix. Shown are the plots for $\gamma=-1.35$ (left, Parallels region) and for $\gamma=-0.9$ (right, North Pole), both at $\omega_{A}=1$ and $j=1$. The shape on the right remains fixed throughout the whole North Pole, and is distorted again only after the separatrix is crossed. Darker regions represent lower levels.}
\end{figure}

\begin{figure}[h]
\scalebox{1.0}{\includegraphics{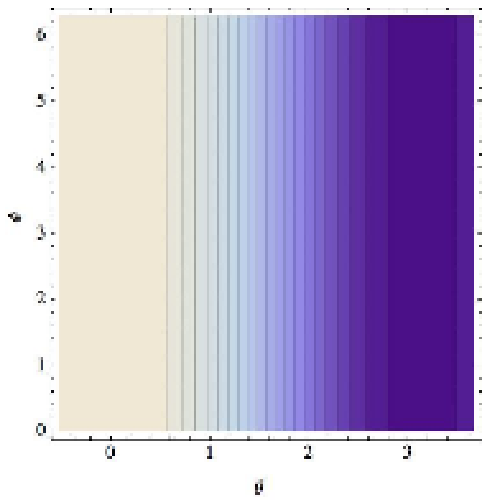}} \qquad
\scalebox{1.0}{\includegraphics{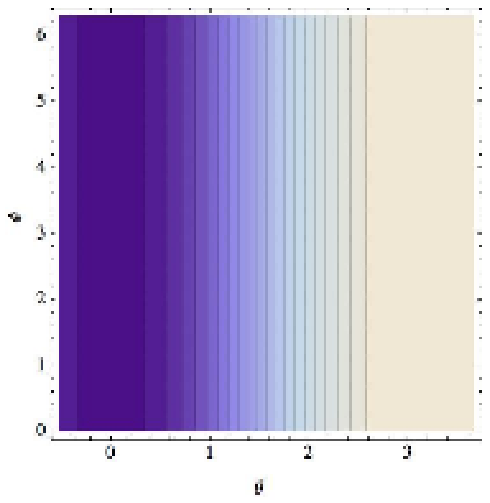}}
\caption{\label{ContourV} Contour levels of the surface energy function $E(\theta,\phi)$ near the vertex of the separatrix. Shown are plots for $\gamma=\omega_{A}=-0.001$ (left, South Pole) and for $\gamma=\omega_{A}=0.001$ (right, North Pole). The change is abrupt. Darker regions represent lower energy levels.}
\end{figure}

The crossing at the vertex $\gamma=\omega=0$ along a non-zero slope, such as Path V in Fig.(\ref{separa}), is a first order phase transition. The minimum energy state switches from all the atoms in their ground state and zero photons in the field (North Pole), to all the atoms in their hyperfine excited state and zero photons in the field (South Pole). This transition involves a change in sign for $\omega_{A}$, which may be achieved by continuously varying a magnetic field to which the atoms in the cavity are exposed. The energy surface $E(\theta,\phi)$ changes abruptly through this vertex, as expected, and it is shown in Fig.(\ref{ContourV}).

\subsection{Expectation Values of Field and Matter Observables }

As mentioned in Section A, the trial state is a tensor product of the coherent states of the one mode electromagnetic field times the collective atomic state. The expectation values of the quadratures of the field, $\hat q,\,\hat p$, and their fluctuations, can therefore be written in terms of real variables $q$ and $p$. In a similar form, we can determine the expectation values of the photon number operator and its corresponding fluctuations. For the matter observables we have a similar situation, and the expectation values of the occupation operator $J_z$ and atom dipole moments $J_x$ and $J_y$ can be expressed in terms of the stereographic projection variables $(\theta, \phi)$ indicating a point on the Bloch sphere.

\begin{table}
\caption{Expectation values of the quadratures of the field, photon number operator and their corresponding fluctuations, at the minima.}
\begin{center}
\begin{tabular}{|c||c|c|c|c|}
\hline
Minima & $\langle \hat q \rangle/\sqrt{N}$ & $\langle \hat p \rangle/\sqrt N$  & $\langle \hat n \rangle/N$ &  $(\Delta \hat n)^2/N^2 $\\
\hline\hline
$\theta_{c}=0$ &\rule[-4mm]{0mm}{10mm}$0$&$0$&0 &0 \\
\hline
$\theta_{c}=\pi$&\rule[-4mm]{0mm}{10mm}$0$
& $0 $  & 0 & 0 \\
\hline
$\theta_{c}=\arccos\left(\omega_{a}/\gamma^{2}\right)$&
\rule[-4mm]{0mm}{10mm}$-\frac{\gamma}{\sqrt2} \Bigl( 1-\frac{\omega_A^2}{\gamma^4}\Bigr)^{1/2}\cos\phi$ &\quad
$\frac{\gamma}{\sqrt2} \Bigl( 1-\frac{\omega_A^2}{\gamma^4}\Bigr)^{1/2}\sin\phi$ & $\frac{\gamma^2}{4} \Bigl( 1-\frac{\omega_A^2}{\gamma^4}\Bigr)$  & $\frac{\gamma^2}{4 \, N} \Bigl( 1-\frac{\omega_A^2}{\gamma^4}\Bigr)$ \\
\hline
\end{tabular}
\end{center}
\label{cfield}
\end{table}

Then the expectation values of the quadratures of the electromagnetic field at the mimimum critical points, given in Eqs.(\ref{quadratureq},\ref{quadraturep}), are shown in Table (\ref{cfield}), where we have divided the expectation values by the appropiate normalization quantities. The fluctuations $\Delta \hat p$ and $\Delta \hat q$  are those associated to the coherent state of the electromagnetic field so they have their minimum values allowed by the Heisenberg uncertainty relations, i.e., $\Delta \hat p = \Delta \hat q = 1/ \sqrt 2$. The expectation value of the number of photons and its corresponding fluctuation at the minimum critical points are also indicated in  Table (\ref{cfield}).  

The expectation values for $J_x$, $J_y$, and $J_z$ per particle are given by the spherical coordinates of a point on a sphere of radius $1/2$. Therefore at the North and South Poles we get $\langle J_x \rangle = \langle J_y \rangle =0$, and $\langle J_z \rangle/N = \pm 1/2$, respectively. For the Parallel region we have
	\begin{equation}
	\frac{\langle J_x \rangle}{N} = \frac{1}{2} \sqrt{1- \frac{\omega_A^2}{\gamma^4}} \cos\phi, \quad  
	\frac{\langle J_y \rangle}{N} = \frac{1}{2} \sqrt{1- \frac{\omega_A^2}{\gamma^4}} \sin\phi , \quad 
	\frac{\langle J_z \rangle}{N} = -\frac{\omega_A}{2 \, \gamma^2}\ .
	\end{equation}
The corresponding fluctuations per particle are indicated in Table (\ref{cmatter}), their values tend to zero as the number of atoms in the system increases.   

\begin{table}
\caption{Fluctuations of the operators $J_x$, $J_y$ and $J_z$ divided by the square of the number of atoms in the system, at the minima.}
\begin{center}
\begin{tabular}{|c||c|c|c|}
\hline
Minima & $ (\Delta J_x)^2/N^2 $ & $ (\Delta J_y)^2/N^2$  & $(\Delta J_z)^2/N^2$ \\
\hline\hline
$\theta_{c}=0 $ &\rule[-4mm]{0mm}{10mm}$\frac{1}{4 N}$&$\frac{1}{4 N}$&0  \\
\hline
$\theta_{c}=\pi$&\rule[-4mm]{0mm}{10mm}$\frac{1}{4 N}$
& $\frac{1}{4 N} $  & 0  \\
\hline
$\theta_{c}=\arccos\left(\omega_{a}/\gamma^{2}\right)$&
\rule[-4mm]{0mm}{10mm}$\frac{1}{4 N} \Bigl( 1-\frac{\omega_A^2}{\gamma^4}\Bigr)\cos^2\phi$ &\quad
$\frac{1}{4 N} \Bigl( 1-\frac{\omega_A^2}{\gamma^4}\Bigr)\sin^2\phi$ & $\frac{1}{4 N} \Bigl( 1-\frac{\omega_A^2}{\gamma^4}\Bigr)$  \\
\hline
\end{tabular}
\end{center}
\label{cmatter}
\end{table}

\subsubsection*{Squeezing and Entanglement Entropy $S_{E}$}

It is well known that coherent states for the electromagnetic field minimize the uncertainty relations. Radiation is said to be squeezed if the uncertainty of one quadrature is less than the standard quantum limit~\cite{dodonov,walls}. For spin or angular momentum systems, ever since the contribution by Kitagawa and Ueda~\cite{ueda} it has been recognized that the phenomena of squeezing is present when the fluctuations of the spin are correlated. It is then straightforward to prove that the $SU(2)$ coherent states or collective atomic states are not squeezed and they have a squeezing parameter equal to unity.  As a consequence, the squeezing parameter
$$
\xi = \sqrt{\frac{2(\Delta J_{\bot})^{2}}{j}},
$$
where $J_{\bot}$ denotes a component of $J$ orthogonal to $\langle J \rangle$, gives us information of how good the trial state is approximating the exact solution of the eigenvalue problem of the TC Hamiltonian, i.e., the approximation is as good as how close the squeezing parameter is to one.

Taking the partial trace for our trial state in the electromagnetic or
in the atomic parts, there is no entanglement entropy. However, there is another way to separate the system and take the partial trace, by considering the Jordan-Schwinger map
	\begin{equation}
		J_{z}=\frac{1}{2}\left(b_{2}^{\dagger}\,b_{2}
		-b_{1}^{\dagger}\,b_{1}\right)\, ,\quad
		J_{+}=b_{2}^{\dagger}\,b_{1}\, ,\quad
		J_{-}=b_{1}^{\dagger}\,b_{2}\ ,
	\end{equation}
where the bosonic operators $b_{i},\,b_{j}^{\dagger}$, $i,\,j=1,\,2$,
satisfy the commutator algebra of two independent harmonic oscillators. Consider taking the partial trace over the electromagnetic and the bosonic operators associated to the excited two level system; the result will be the same as if we take the trace over the radiation field and over the ground state level of the density matrix of the coherent atomic state~\cite{milburn}. 

The coherent atomic state in terms of states of a two dimensional harmonic oscillator can be written as
	\[
		\vert\zeta\rangle=\sum_{n_{2}=0}^{2\,j}\binom{2j}{n_2}^{1/2}\,\frac{\zeta^{n_{2}}}{\left(1+
		\left|\zeta\right|^{2}\right)^{j}}\,\vert 2\,j-n_{2},\,n_{2}\rangle\ ,
	\]
where the state $\vert 2 j -n_2, n_2 \rangle \equiv \vert n_1,n_2 \rangle$, with $n_1$ denoting the number of $b_1$ bosons (number of atoms in the lowest state) and $n_2$ the number of $b_2$ bosons (number of atoms in excited level), with the constraint that the total number of quanta $n_1 + n_2$ must equal the total number of atoms $N= 2 j$ in the system.
 
The partial trace over the first oscillator of the density matrix
associated to this atomic coherent state is
	\begin{equation}
		\varrho_{\zeta}^{\ 2}=\hbox{Tr}_{1}\left(\varrho_{\zeta}\right)=\sum_{n=0}^{N}
		\binom{N}{n}\,\frac{\left|
		\zeta\right|^{2\,n}}{\left(1+\left|\zeta\right|^{2}
		\right)^{N}}\,\vert n\rangle\langle n\vert\ .
		\label{rhoc}
	\end{equation}
Therefore, the reduced density matrix is diagonal. Using Eq.(\ref{dzeta}) the entanglement entropy is
	\begin{equation}
		S_{E}(\theta)=-\sum_{n=0}^{N}\,p_{n}(\theta)\,\ln(p_{n}(\theta))\ ,
		\label{coherententropy}
	\end{equation}
where
	\[
		p_{n}(\theta)=\binom{N}{n}\,
	\biggl( \frac{1 - \cos\theta}{2}\biggr)^{n}	
	\biggl( \frac{1 +\cos\theta}{2} \biggr)^{N-n}\ ,
	\]
is the probability of a binomial distribution, which also corresponds to the occupation probabilty of $n$ particles in the excited state of a two-level system. Through this expression it is immediate to determine the estimates on the occupation probability and the entanglement entropy of the ground state. The results are given in Table (\ref{sqent}).

\begin{table}
\caption{Occupation probability and entanglement entropy of the ground state of the TC model by means of the tensorial product of coherent states as trial state.}
\begin{center}
\begin{tabular}{|c||c|c|}
\hline
Minima & $ p_n $ & $ S_E $    \\
\hline\hline
$\theta_{c}=0 $ &\rule[-4mm]{0mm}{10mm}$\delta_{n,0}$&$ 0 $ \\
\hline
$\theta_{c}=\pi$&\rule[-4mm]{0mm}{10mm}$\delta_{n,N}$
& $0 $   \\
\hline
$\theta_{c}=\arccos\left(\omega_{a}/\gamma^{2}\right)$&
\rule[-4mm]{0mm}{10mm}$\quad \binom{N}{n} \bigl( \frac{1 - {\omega_A}/{\gamma^2}}{2}\bigr)^{n} \bigl( \frac{1 + {\omega_A}/{\gamma^2}}{2} \bigr)^{N-n} $ \quad &\quad
$-\sum_{n=0}^{N}\,p_{n}\,\ln(p_{n})$ \quad  \\
\hline
\end{tabular}
\end{center}
\label{sqent}
\end{table}

\subsection{Trial state in control parameter space}

The expression (\ref{trial}) for the trial state $\vert \alpha,\, \zeta \rangle$ that minimizes the energy surface takes the following forms: At the North Pole, $\omega_{A}>\gamma^{2}$, the ground state is given by  
	\begin{equation}
		\vert\psi_{np}\rangle=\vert 0\rangle\otimes\vert j,\,-j\rangle\ .
	\end{equation}
At the South Pole, $\omega_{A}<-\gamma^{2}$, the trial state has the form  
	\begin{equation}
		\vert\psi_{sp}\rangle=\vert 0\rangle\otimes\vert j,\, j\rangle\ .
	\end{equation}
For the Parallel case, $\left|\omega_{A}\right|<\gamma^{2}$, the approximate expression to the ground state is given by
	\begin{equation}
		\vert\psi_{par} \rangle=\sum_{m=-j}^{+j}\ \sum_{\nu=0}^{+\infty}\,
		A_{m,\,\nu} \vert\nu\rangle\otimes\vert j,\,m\rangle\ ,
	\end{equation}
where we define the expansion coefficients
	\begin{eqnarray}
		A_{m,\,\nu} &=&\binom{2j}{j+m}^{1/2}\, \exp\left\{-\frac{j\, \gamma^{2}}{4}\left(1-\frac{
		\omega_{A}^{2}}{\gamma^{4}}\right) + i\,(j+m-\nu)\,\phi \,\right\}
		\nonumber \\
		&&\times \frac{\left(-\sqrt{2j}
		\,\gamma\right)^{\nu}}{\sqrt{\nu!}}
		\left(\frac{1}{2}+\frac{\omega_{A}}{2\,\gamma^{2}}\right)^{(j-m+\nu)/2}\,
		\left(\frac{1}{2}-\frac{\omega_{A}}{2\,\gamma^{2}}\right)^{(j+m+\nu)/2}\, .
	\label{f1}
	\end{eqnarray}
One usually would perform this sum by selecting a value of $\nu$ to make the sum over
$m$, and then proceeding with the following value of $\nu$, until we
reach some type of convergence. In the TC model $\lambda=m+\nu$ is a
conserved quantity. By replacing $m$ by $\lambda-\nu$, we can write
	\begin{equation}
		\vert\psi_{par} \rangle=\sum_{\lambda=-j}^{+j}\ \sum_{\nu=0}^{\lambda+j}
		A_{\lambda-\nu,\,\nu} \vert\nu\rangle \otimes \vert j,\,\lambda-\nu\rangle\
		+\sum_{\lambda=j+1}^{+\infty} \\ 
		\sum_{\nu=\lambda-j}^{\lambda+j}A_{\lambda-\nu,\,\nu} 
		\vert\nu\rangle \otimes \vert j,\, \lambda-\nu\rangle\ ,
	\end{equation}
which, in compact form, is
	\begin{equation}
		\vert\psi_{par} \rangle=\sum_{\lambda=-j}^{+\infty}
		\sum_{\phantom{of}\nu=\max(0,\lambda-j)}^{\lambda+j}
		A_{\lambda-\nu,\,\nu}\ \vert\nu\rangle \otimes \vert j,\, \lambda-\nu\rangle\ .
	\end{equation}

\section{Quantum Analysis}

\subsection{Ground State Energies and Constant of Motion}

To understand the physical meaning of the constant of motion
$\Lambda=a^{\dagger}a+J_{z}$, we apply the transformation
	\begin{equation}
		{\cal U}(\eta)=\exp\left(i\,\eta\,\Lambda\right)\ .
	\end{equation}
to the TC Hamiltonian. The effect of the transformation in the   
quadratures of the electromagnetic field is
	\begin{equation}
		{\cal U}(\eta)\,\left(\begin{array}{c}\hat{q}\\\hat{p}\end{array}
		\right)\,{\cal U}^{\dagger}(\eta)=\left(\begin{array}{cc}
		\cos\eta&\sin\eta\\-\sin\eta&\cos\eta\end{array}\right)\,
		\left(\begin{array}{c}\hat{q}\\\hat{p}\end{array}
		\right)\ ,
	\end{equation}
making a rotation by an angle $\eta$
counterclockwise in the plane of the electromagnetic quadratures.
In the same way we can calculate the induced transformation for the matter 
observables $J_x$ and $J_y$
	\begin{equation}
		{\cal U}(\eta)\,\left(\begin{array}{c}J_{x}\\J_{y}\end{array}
		\right)\,{\cal U}^{\dagger}(\eta)=\left(\begin{array}{cc}
		\cos\eta&-\sin\eta\\\sin\eta&\cos\eta\end{array}\right)\,
		\left(\begin{array}{c}J_{x}\\J_{y}\end{array}
		\right)\ ,
	\end{equation}
which is a rotation by an angle $\eta$ clockwise along the $z$-axis. This
transformation then leaves invariant each term in the TC
Hamiltonian. Therefore, we propose that the quantum states for the TC Hamiltonian have the form 
	\begin{equation}
		\vert\psi_{k}\rangle=\sum_{\nu=\max[0,\lambda-j]}^{\lambda+j}\,c^{(k)}_{\nu}
		\,\vert\nu\rangle\otimes\vert j,\,\lambda-\nu\rangle\ ,
		\label{quantum-gs}
	\end{equation}
because only one value of $\lambda$ is present. This contrasts with the trial
function that we are considering, which is a superposition of all $\lambda$'s. The index $k$ is denoting the different eigenvectors of the Hamiltonian matrix, whose dimension depends on the value of $\lambda$: for $\lambda \leq j$ one has matrices of dimension $d= \lambda + j + 1$, while if $\lambda \geq j$ the matrices have dimensions $d= 2 j + 1$. We are using the natural basis states to diagonalize the TC Hamiltonian because it has the constant of motion $\Lambda$; for the general case, when one is not considering the rotating wave approximation, a new technique has been proposed to obtain the exact result for systems with size two orders of magnitude higher than reported in the literature~\cite{chen}.

In this contribution we will mainly be concerned with the behaviour of the ground state, which will be denoted  by $\vert \psi_{gs} \rangle$ and its expansion coefficient by $c_\nu$.

Finding analytic solutions for small values of $\lambda$ is straightforward, for an arbitrary number of particles $N$, any value of the detuning parameter $\Delta$, and an arbitrary coupling $\gamma$:
\begin{enumerate}
\item[i)]
For $\lambda = - \frac{N}{2}$ we have a $1 \times 1$ Hamiltonian matrix, the ground state energy is given by $E_{0}/N = -\frac{1}{2}(1-\Delta),$
and the state is $\vert 0\rangle\otimes\vert j,\, -j\rangle$.
\item[ii)]
For $\lambda = - \frac{N}{2} + 1$ we have a $2 \times 2$ Hamiltonian matrix, the ground state energy is given by $E_{1}/N = \frac{1}{2 N} (2 - N - \Delta + N\Delta - \sqrt{4 \gamma^2 + \Delta^2}),$
and the state is a linear combination involving $0$ and $1$ photons: $\vert 0\rangle\otimes\vert j,\, -j+1\rangle$ and $\vert 1\rangle\otimes\vert j,\, -j\rangle$.
\item[iii)]
For $\lambda = - \frac{N}{2} + 2$ we have a $3 \times 3$ Hamiltonian matrix, the ground state energy is given by
$$E_{2}/N = \frac{N \Delta - N - 2 \Delta +4}{2
   N}-\frac{2 \sqrt{\frac{(4 N-2)
   \gamma ^2}{N}+\Delta ^2}\, \sin
   \left(\frac{1}{6} (2 \phi +\pi )\right)}{\sqrt{3}
   N},$$
where $N \geq 2$ and
$$\phi = \tan ^{-1}\left(\frac{N^2
   \sqrt{\left(\frac{(4 N-2) \gamma
   ^2}{N}+\frac{\Delta
   ^2}{N^2}\right)^3-\frac{27 \gamma ^4
   \Delta ^2}{N^4}}}{3\, \sqrt{3}\, \gamma ^2
   \Delta }\right),$$
and the state is a linear combination involving $0$, $1$ and $2$ photons: $\vert 0\rangle\otimes\vert j,\, -j+2\rangle$, $\vert 1\rangle\otimes\vert j,\, -j+1\rangle$ and $\vert 2\rangle\otimes\vert j,\, -j\rangle$.
\end{enumerate}

The expressions get more and more complicated, but one can solve analytically for up to $\lambda = - \frac{N}{2} + 4$. The way in which our Hamiltonian is written allows us to take the thermodynamic limit $N \rightarrow \infty$ in the expressions for the energy; in all cases mentioned these reduce to $E_0/N$.  When in resonance $\Delta=0$, the results for all these values of $\lambda$ coincide with those given by Bu\v{z}ek et al.~\cite{buzek}.

\begin{figure}[h]
\scalebox{1.2}{\includegraphics{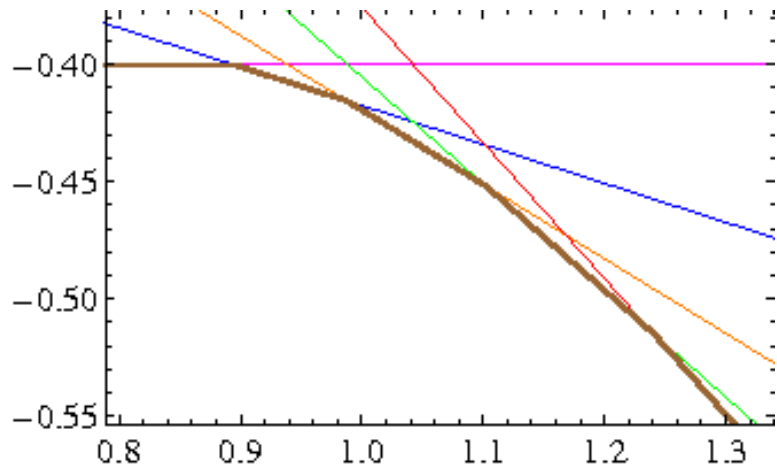}} 
\caption{\label{new_energy_spectrum} Ground state energy as a function of coupling constant $\gamma$. Results are shown for values of $\lambda$ from -3 to +1, for 6 atoms and $\Delta=0.2$. The heavy line represents the minimum energy for all $\gamma$, and each change in $\lambda$ reflects a quantum phase transition.}
\end{figure}

In Fig.(\ref{new_energy_spectrum}) we show the quantum phase transitions for a system of 6 atoms ($j=3$), with a detuning parameter $\Delta=0.2$. The straight lines correspond to the energy of the ground state for different values of the constant of motion $\lambda$, starting from $-3$ ($0$ photons) to $+1$ ($4$ photons). In each case the eigenstate is a mixture of $0$ to $\lambda+j$ photons.

By taking the trace with respect to the field (matter) states the reduced density matrix takes the form
$$
\varrho^{matter} = \sum_{n=0}^{\min\{\lambda + \frac{N}{2},\,N\}} \left| c_{\lambda + \frac{N}{2} - n} \right|^2 \vert N-n,\,n\rangle \langle N-n,\,n\vert
$$
$$
\varrho^{field} = \sum_{\nu=\max\{0,\lambda - j\}}^{\lambda + j} \left| c_{\nu} \right|^2 \vert \nu\rangle \langle \nu\vert
$$
where $c_{\lambda + \frac{N}{2} - n}$ (or $c_\nu$) is determined from the Hamiltonian diagonalization.  Taking the trace over the first oscillator in $\varrho^{matter}$, we have a diagonal matrix of the same size given by
	\begin{equation}
	\left( \varrho_{0}^{\ 2} \right)_{n_1,n_2} = \left| c_{\lambda + \frac{N}{2} - n_1} \right|^2 \delta_{n_1,n_2}
	\label{rhoq}
	\end{equation}
where $n_1=0,\cdots, \min\{\lambda + \frac{N}{2},\,N\}$.
These expressions will be used below to calculate the fidelity between the variational and the exact quantum states. As the reduced density matrix of the matter is diagonal, the matter-field entanglement entropy equals that between the atoms occupying the two hyperfine levels
$$
S_E = - \sum_{n=0}^{\min\{\lambda + \frac{N}{2},\,N\}} \left| c_{\lambda + \frac{N}{2} - n}\right|^2 \ln \left| c_{\lambda + \frac{N}{2} - n}\right|^2
$$
in both cases.

The expectation values for the photon, $\hat{\cal O_F}$, and atomic, $\hat{\cal O_A}$ observables with respect to the quantum ground state can be simplified to the expressions 
	\begin{eqnarray}
		\langle\psi_{gs}\vert\,\hat{\cal O}_{F}\,\vert\psi_{gs}\rangle
		&=&\sum_{\nu=\max[0,\lambda-j]}^{\lambda+j}\,\left|c_{\nu}\right|^{2}\,
		\langle\nu\vert\,\hat{\cal O}_{F}\,
		\vert\nu\rangle\ \\
		\langle\psi_{gs}\vert\,\hat{\cal O}_{A}\,\vert\psi_{gs}\rangle
		&=&\sum_{\nu=\max[0,\lambda-j]}^{\lambda+j}\,\left|c_{\nu}\right|^{2}\,
		\langle j,\,\lambda-\nu\vert\,\hat{\cal O}_{A}\,
		\vert j,\,\lambda-\nu\rangle\ .
	\end{eqnarray}

Given the form of the quantum ground state (\ref{quantum-gs}) it is
immediate that: 1) the expectation values of any power of the ladder operators
$J_{\pm}$ vanish because $\lambda$ and the number of excited atoms cannot be changed when the number of photons remains constant. Therefore
$\langle J_{x}\rangle=\langle J_{y}\rangle=0$, and $\langle J_{x}^{2}\rangle
=\langle J_{y}^{2}\rangle$ because we have only contributions from the diagonal part of the operators. 2) The expectation values of any power of the creation and annihilation operators vanish for the same reason. In particular we obtain $\langle \hat q\rangle=\langle \hat p\rangle
=\langle\left(\hat q \hat p + \hat p\hat q\right)\rangle=0$. For the expectation values of $\hat q^2$ and $\hat p^2$, one has again contributions from the diagonal part and they are given by
	\[
	\langle \hat q^2 \rangle = \langle \hat p^2 \rangle =  \langle \hat n + 
	\frac{1}{2} \rangle = \nu + \frac{1}{2} \, ,
	\]
where $\nu$ denotes the number of photons of the state. 3) The fluctuations in the quadratures of the fiels are given by $( \Delta \hat p)^2 = \langle \hat q^2 \rangle $ and 
$( \Delta \hat q)^2 = \langle \hat p^2 \rangle $.

As in the TC Hamiltonian the expectation value of the angular momentum vector is only in direction $z$, and the fluctuations of the dipole moment components of the atoms satisfy $(\Delta J_x)^2 = (\Delta J_y)^2$; the squeezing spin coefficients in directions $x$ and $y$ must be equal. Therefore, the expression for $\xi$, in the orthogonal directions to the $z$-axis, is defined by the expression
	\begin{equation}
		\xi_{k}=\sqrt{\frac{2(\Delta J_{k})^{2}}{j}}=\sqrt{j+1-\frac{\langle
		J_{z}^{2}\rangle}{j}}\ ,
	\end{equation}
where $k=x$ or $y$.

\section{Semiclassical versus Quantum Results}

\subsection{Ground State Energies and Constant of Motion Expectation Values}

\begin{figure}[h]
\scalebox{0.7}{\includegraphics{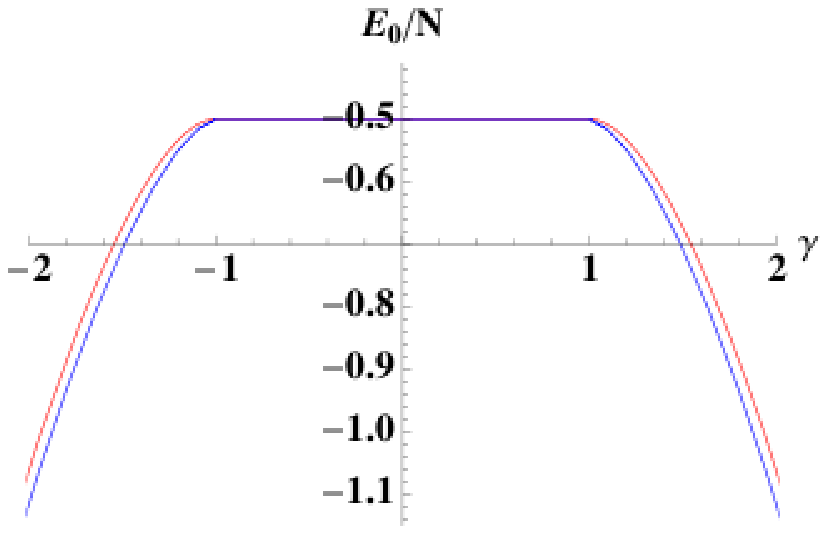}} \quad
\scalebox{0.7}{\includegraphics{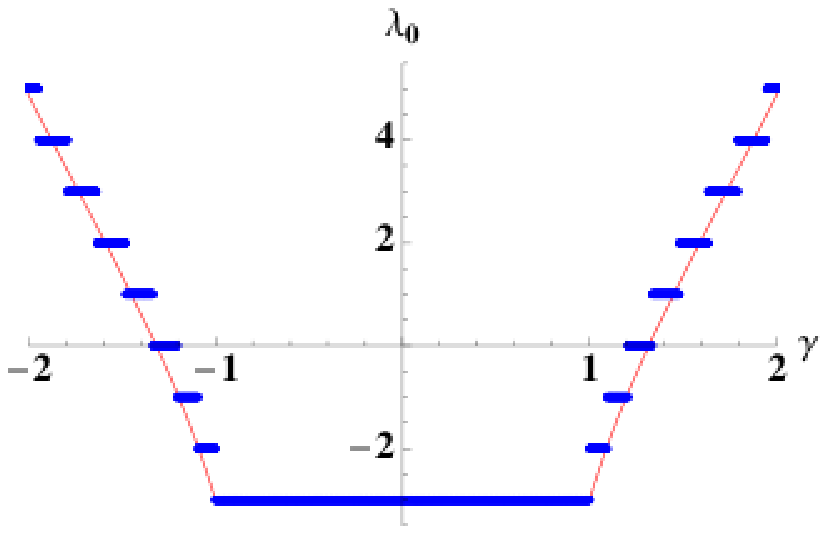}}

\

\scalebox{0.7}{\includegraphics{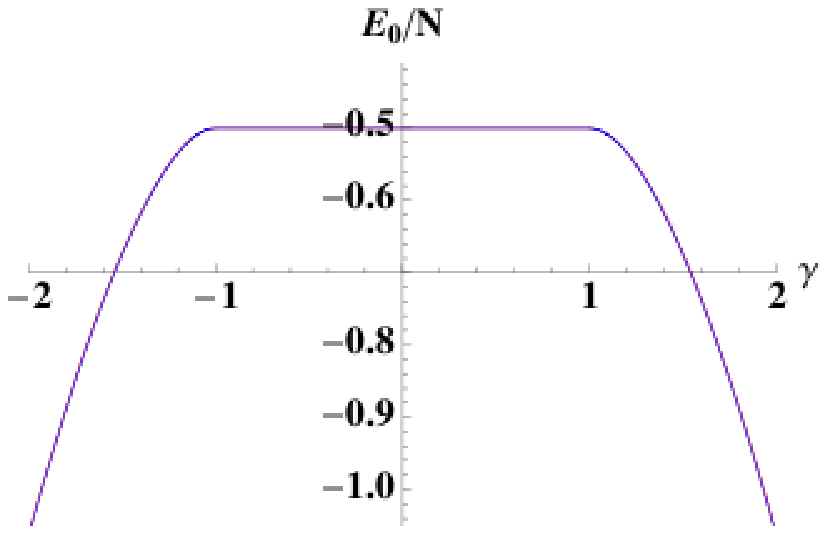}} \quad
\scalebox{0.5}{\includegraphics{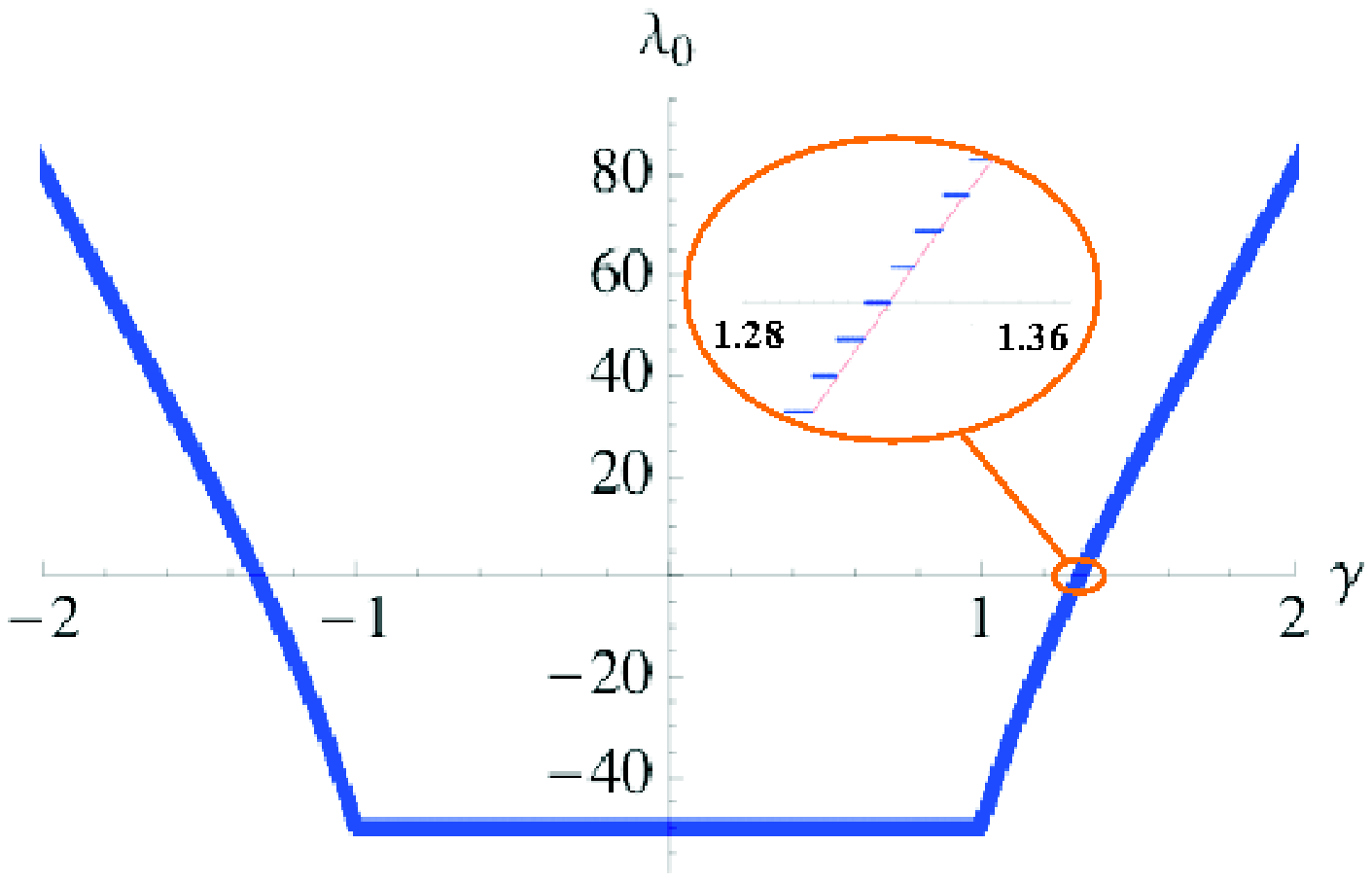}}
\caption{\label{energy} The energy per number of atoms, $E_0/N$ (left), and the constant of motion, $\lambda_{0}$ (right), are shown as functions of the interaction parameter $\gamma$ for both, the semiclassical and quantum cases. The exact quantum energy is equal to or less than its classical counterpart. The semiclassical constant $\lambda_{0}$ has continuous values while the exact quantum result is discrete. We use $N=6$ (top), and $N=100$ (bottom). In both cases $\Delta=0$. For $N=100$ the graphs totally superimpose; however, the zoom in the last graph makes the discreteness in $\lambda_{0}$ evident.}
\end{figure}

Fig.(\ref{energy}) displays the energy per number of particles $E_0/N$, and the constant of motion $\lambda_{0}$, as a function of the coupling interaction $\gamma$. The top figures correspond to a number of atoms $N=6$, while the bottom figures correspond to $N=100$; in both cases the detuning parameter is $\Delta=0$. While the classical energy slightly overestimates the exact quantum result outside the North Pole (as must be according to the Rayleigh-Ritz variational principle), second order corrections to the eigenvalues of the Hamiltonian matrix gives a much better agreement when the number of atoms is increased, as shown.

 We observe that both $E_{0}$ and $\lambda_{0}$ depend strongly on the coupling strength $\gamma$. As this coupling is increased, the value of the (quantum) constant of motion at which the minimum energy is obtained remains constant until a threshold is reached, at which point it jumps to its next value. This same process repeats itself as $\gamma$ keeps increasing, $\lambda_{0}$ taking discrete values throughout (in contrast to its semiclassical counterpart, which is continuous), reflecting the quantum nature of the system. As $N$ increases, the width of the steps becomes smaller until the difference with the classical result seems negligible. However, the separation of $\lambda$-values always equals $1$ (cf. zoom in the figure). In the North Pole, the state is $\vert\nu\rangle\otimes\vert j,\,m\rangle = \vert 0\rangle\otimes\vert j,\,-j\rangle$, so only one value of $\lambda_{0}$ (and of $E_{0}$) is had all through this region $-1<\gamma<1$, and it naturally coincides with the classical value.  The same (dual) result is obtained in the South Pole.
 
This discrete behaviour in $\lambda_{0}$ is inherited in all the quantum observables of interest, as we shall see below. It is important to note the horizontal shape of the steps in $\lambda_{0}$ shown in Fig.(\ref{energy}).  This is a consequence of the system being in resonance: $\Delta=0$.  When the detuning parameter is taken away from resonance, the $\lambda_{0}$ steps acquire a slope given precisely by the difference in energy between the absorbed photons and the atom's energy level separation.

\begin{figure}[h]
\scalebox{1.2}{\includegraphics{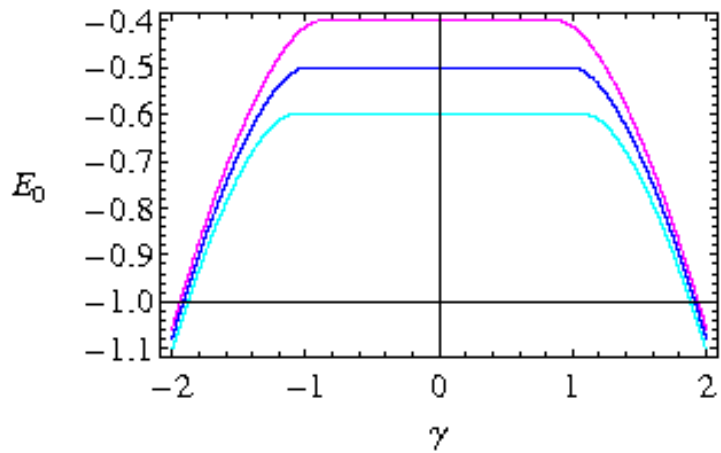}}
\caption{\label{delta} Energy per number of atoms, $E_0/N$ as a function of the parameter $\gamma$. We use $j=10$. The lowest energy curve corresponds to $\Delta=-0.2$, the intermediate energy curve to $\Delta=0$, and the curve with larger values to $\Delta=0.2$.}
\end{figure}

Given the present possibility of confining a small number of atoms and performing non-demolition experiments on them, and the direct importance of these on cavity QED, quantum information theory, encryption, teleportation, and quantum optics in general, in all the results that follow, we take $N=6$ and $\Delta=0.2$ in order to illustrate the behaviour of a small system slightly away from resonance. The effect of the detuning parameter on the ground state energy, for $\Delta=\ -0.2,\,0,\,0.2$, is shown in Fig.(\ref{delta}). One notices that the width of the North Pole depends on the value of $\Delta$, as expected.

\subsection{Matter Observables}

\begin{figure}[h]
\scalebox{0.75}{\includegraphics{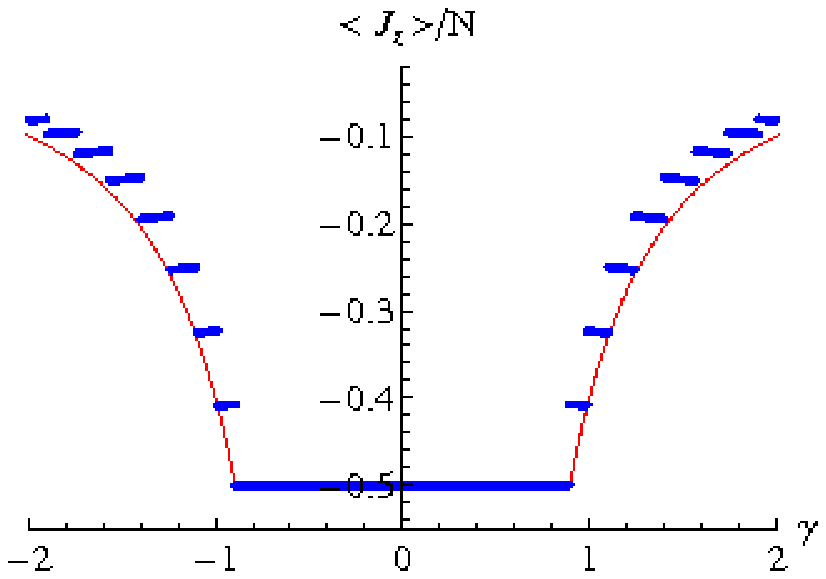}}
\scalebox{0.75}{\includegraphics{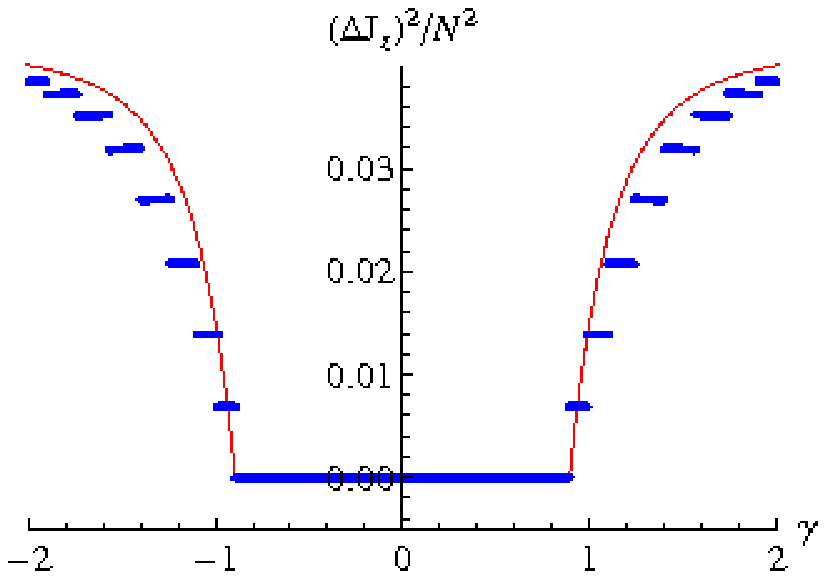}}
\caption{\label{jzeta} The expectation value $\langle J_z\rangle/N$ (left) for the semiclassical (continuous) and quantum (discrete) models, together with their corresponding fluctuations (right), are displayed as functions of $\gamma$, for $N=6$ and $\Delta=0.2$.}
\end{figure}

Fig(\ref{jzeta}) presents the expectation value of $\langle J_z \rangle/N$ for $N=6$ atoms and $\Delta=0.2$. The observable takes discrete values for the exact quantum solution while it is continuous for the semiclassical trial state. The inherited discreteness from $\lambda_{0}$, associated to quantum phase transitions as the structure of the ground state changes with $\gamma$, is evident.  The semiclassical approximation follows the same trend.

For $\gamma^2 < \omega_A$ all the atoms are concentrated in their lowest hyperfine level. It is natural to find that $\langle J_z\rangle/N$ takes on the value $-j/N$, which in the plotted case corresponds to 
$-0.5$, and the constant of motion takes its minimum value $\lambda=-3$.  For values of $\gamma$ outside this region there are other occupancies. When $\gamma$ is very large the occupancy tends to zero, i.e., the two hyperfine levels are equally occupied. 
The dispersion $(\Delta J_z)^2$ has values ten times smaller.

\subsection{Field Observables}

\begin{figure}[h]
\scalebox{0.75}{\includegraphics{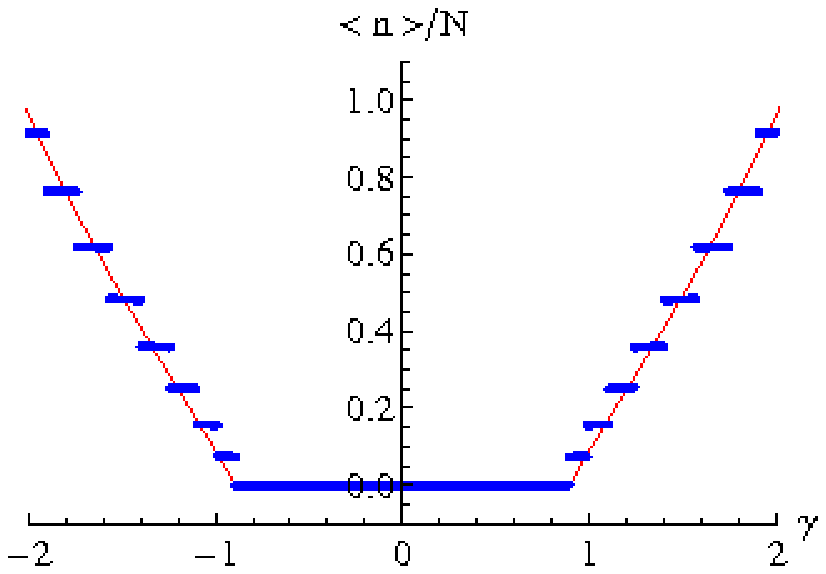}}
\scalebox{0.75}{\includegraphics{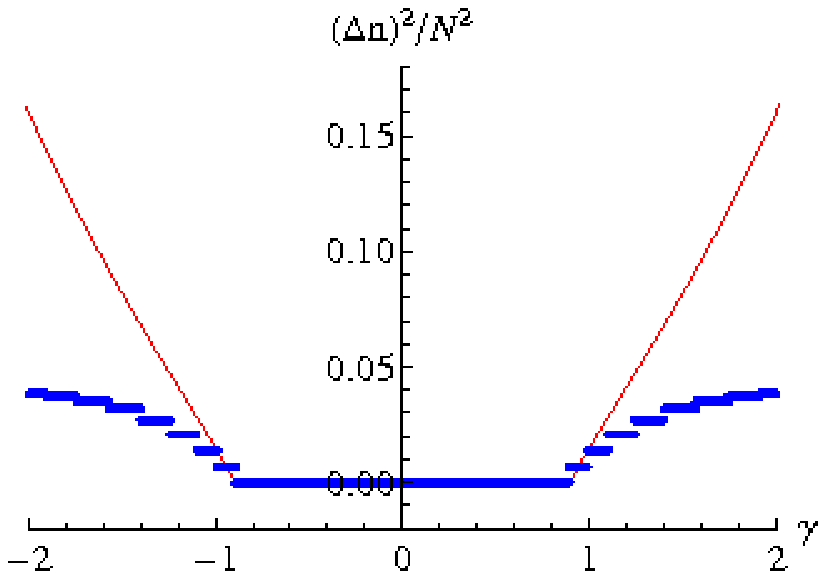}}
\caption{\label{nfoton} The expectation value of $\langle \hat n \rangle/N$ (left) and its corresponding fluctuation $( \Delta n)^2/N^2$ (right), are shown as functions of $\gamma$. We use $N=6$, and $\Delta=0.2$.}
\end{figure}

The expectation value $\langle \hat n \rangle/N$ of the photon number operator per number of atoms with respect to the ground state is shown in Fig.(\ref{nfoton}), for the exact and variational quantum results, for a system of $N=6$ atoms with a detuning $\Delta=0.2$. The fluctuation in the number of photons $( \Delta n)^2/N^2$ is also shown. Note that in the North Pole, $\vert\nu\rangle\otimes\vert j,\,m\rangle = \vert 0\rangle\otimes\vert j,\,-j\rangle$, the number of photons is zero. Thus, the fluctuation {\it must} be zero and completely coincide with its semiclassical counterpart.  Outside this region, the proposed trial function as a coherent state for the description of the electromagnetic field, which includes an infinite number of photons in its composition, greatly overestimates the quantum result: for $\gamma\approx 2$ there is approximately one photon per atom with very good agreement between the two procedures, while their corresponding fluctuations have at least a difference of a factor of six.

\subsection{Squeezing, Entanglement and Fidelity}

\begin{figure}[h]
\scalebox{0.75}{\includegraphics{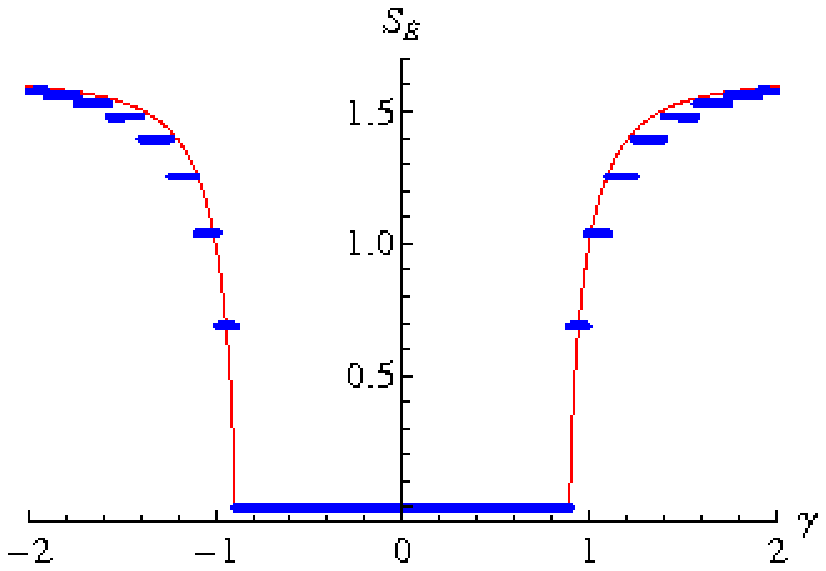}}
\scalebox{0.75}{\includegraphics{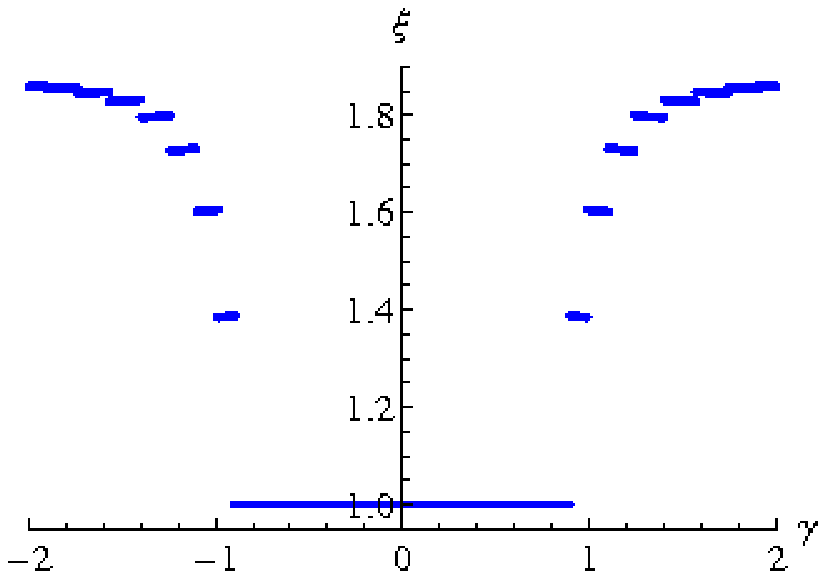}}
\caption{\label{sqe} The entanglement entropy (left) and the squeezing spin coefficient (right) are plotted as functions of $\gamma$, for $N=6$, and $\Delta=0.2$.}
\end{figure}

We have found that the ground state of the TC model does not present the phenomenon of squeezing in the matter components. However we propose to use the behaviour of the squeezing spin coefficient $\xi$ as an additional signature of the goodness of our trial state in reproducing the exact quantum solution. We can see in Fig.(\ref{sqe}) that the squeezing coefficient exceeds the value of one for large values of the coupling parameter. This means that the trial state must have a very small overlap with the exact quantum state, and it is due to the multiple values of the constant of motion contained in the composition of our trial state.
 
The entanglement is usually regarded as a purely quantum correlation which plays an important role in quantum phase transitions~\cite{lambert, buzek}. For the variational state the entanglement entropy between the matter and field is zero because it corresponds to a pure state (matter coherent state). It is then interesting to do a comparison with the entanglement between atoms in the semi-classical approach. Fig.(\ref{sqe}) shows the entanglement entropy associated to the variational and exact quantum cases, for a reduced density matrix that eliminates the electromagnetic degrees of freedom and discriminates the matter part into the atoms occupying the lowest hyperfine level from the rest, and viceversa.  Once again, the semiclassical result is a very good approximation to the exact one. The plots shown are for $N=6$ and $\Delta=0.2$. It is clear that for any $\lambda > -j$ (i.e., outside $\left|\gamma\right|\leq\sqrt{1-\Delta}$) there is entanglement between the atoms. In the thermodynamic limit $N \rightarrow \infty$, the behavior is the same with an even more abrupt slope at the separatrix. Besides, one can notice directly from this plot, that the entanglement entropy with $\gamma\approx 2$ is $S_E \approx 1.6$ nats; this compares favorably with a straightforward estimate of the maximum entanglement entropy available for a quantum system of $j=3$, which would yield $S_E =1.94$ nats.

An interesting measure, coming from quantum information theory, is the fidelity $F(\rho_1,\,\rho_2)$~\cite{jozsa94}, where $\rho_i$ denotes the density matrix of a system. This is a measure of how close two probability distributions are to each other, even for mixed states. In the case of two pure states, the measure is equivalent to the overlap.  In our case, we want to compare the reduced probability distributions calculated from the exact model and from the variational model.  Since they are both diagonal (cf. Eqs.(\ref{rhoc}, \ref{rhoq})), the expression for the fidelity is simple:
	\begin{equation}
		F(\varrho_{\zeta}^{\ 2},\varrho_{0}^{\ 2}) = \sum_{n=0}^{\min\{\lambda + \frac{N}{2}, N\}} \sqrt{\vert c_		{\lambda + \frac{N}{2} - n}\vert^2\, p_n(\theta_c)}\ .
	\end{equation}
	
\begin{figure}[h]
\scalebox{1.1}{\includegraphics{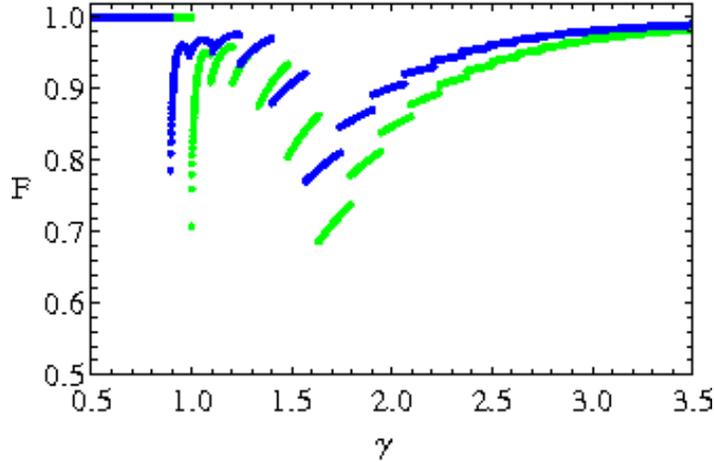}}
\caption{\label{fidelity} Fidelity as a function of $\gamma$, for $j=3$. Light curves correspond to resonance $\Delta=0$, and dark curves to $\Delta=0.2$.}
\end{figure}

Fig.(\ref{fidelity}) shows the fidelity as a function of $\gamma$ both in resonance $\Delta=0$, and away from resonance $\Delta=0.2$, for the probability distributions mentioned above. At the North Pole (zero photons and all atoms in the ground state, $\lambda=-j$) the fidelity represents the overlap between the two states, and equals $1$. Each jump represents a value of $\lambda$ increased by $1$. The first and largest transition is found at the separatrix, while the other discontinuities occur at the quantum phase transitions. Note that $F$ diminishes at each of these for small values of $\lambda$ (small number of photons). In all these cases we have mixed states. After $\lambda=j$ (number of photons equal to the number of atoms) the dimension of the Hamiltonian matrix remains constant and $F$ increases at each transition (except for the anomalous case $N=1$). It is important to stress the fact that the fidelity is high and the probability distribution of the coherent state approximates very well the exact solution, especially for a small number of atoms and a large number of photons.

\bigskip	

\subsection{Occupation Probability Distributions}

\begin{figure}[h]
\scalebox{0.5}{\includegraphics{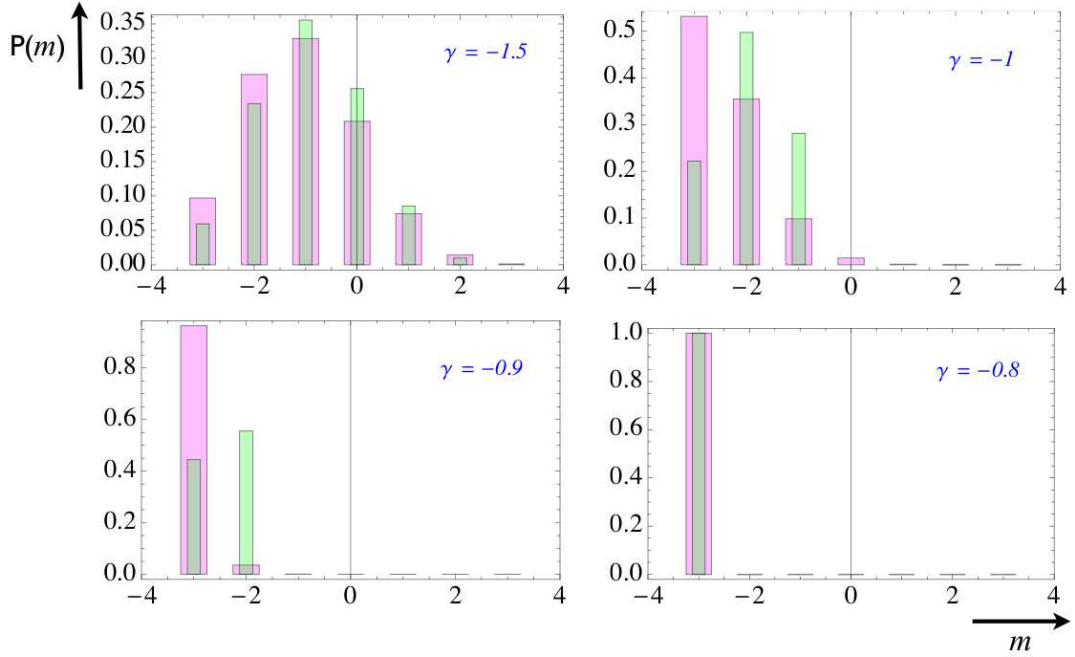}}
\caption{\label{ocprob} The composition of the trial and exact states of the TC model are displayed for $N=6$, $\Delta=0.2$, as we cross the separatrix along $Path\ II$ for values of $\gamma =\, -1.5,\,-1.0,\,-0.9,\,-0.8$. The corresponding quantum eigenvalues of the constant of motion are $\lambda = \,2,\,-1,\,-2,\,-3$. $m$ runs from $-j$ to $j$ ($-3$ to $3$ in this case) along the horizontal axis. The narrow bars correspond to quantum values.}
\end{figure}

\begin{figure}[h]
\scalebox{0.55}{\includegraphics{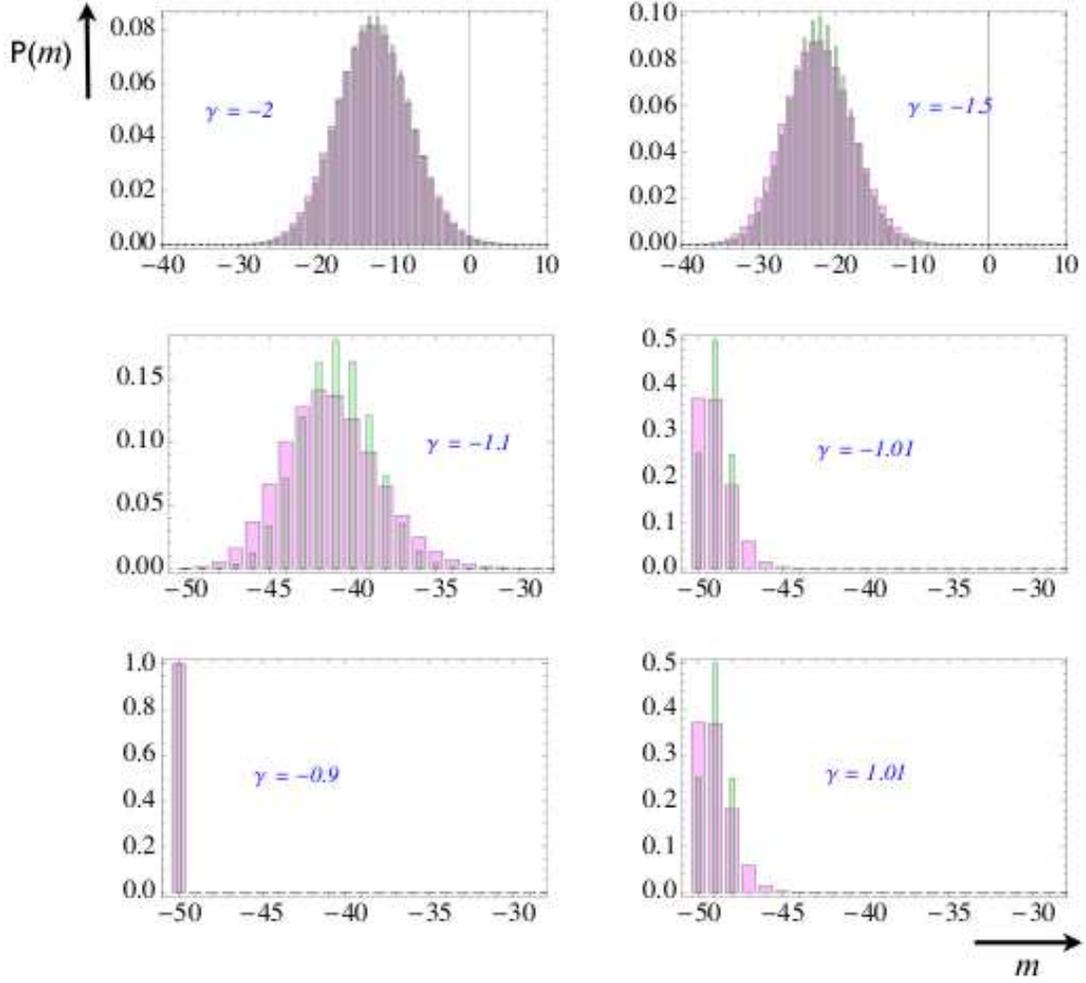}}
\caption{\label{ocprob50} The composition of the trial and exact states of the TC model are displayed for $N=100$, $\Delta=0$, as we cross the separatrix along $Path\ II$ for values of $\gamma =\, -2,\,-1.5,\,-1.1,\,-1.01,\,-0.9,\,1.01$. The corresponding eigenvalues of the constant of motion are $\lambda = \,81,\,23,\,-31,\,-48,\,-50,\,-48$. $m$ runs along the horizontal axis. The narrow bars correspond to quantum values.}
\end{figure}

The composition of the ground state of the TC model can be analyzed both, in terms of the distribution of the atoms into the two hyperfine levels, and of the photon number content, be it for the exact energy ground state or for its variational approximation.  Fig.(\ref{ocprob}) shows the composition of the trial and exact states for $N=6$, $\Delta=0.2$.  The four plots show the change in the occupation probability as one crosses one arm of the separatrix: in this case, we move along $Path\ II$ in Fig.(\ref{separa}) and cross the separatrix at $\gamma=-\sqrt{0.8}$ from the Parallels region into the North Pole (recall that $\Delta=0.2$).  The marked change due to the phase transition is obvious. While the composition of the ground state in both cases are very similar, one has to notice that the variational state is constituted by a distribution of $\lambda$'s in contrast to the exact quantum result which is described by only one $\lambda$-value. (This good approximation is a consequence of the fact that the distribution of $\lambda$'s is a gaussian centered at the quantum eigenvalue, as will be seen below in Fig.(\ref{lams}) and its corresponding discussion.) This quantum $\lambda$-value depends of the coupling parameter $\gamma$ considered in the TC Hamiltonian as can be seen in Fig.(\ref{energy}). It is important to stress that, for values of $\gamma$ far away from the separatrix ($\gamma = -1.5$ in the figure), the semiclassical occupation probability estimates very well the quantum result. The estimate is even better as $N$ increases, until they become almost indistinguishable, as shown in Fig.(\ref{ocprob50}) calculated for $N=100$. However, as we get close to the separatrix ($\gamma=-\sqrt{0.8}$ for $N=6$; $\gamma=-1$ for $N=100$) the semiclassical approximation gets poorer. When we cross it, both distributions become a spike at $\lambda=m=-j$, and they remain so until the separatrix is crossed again. The distributions are exactly the same changing $\gamma$ for $-\gamma$ (cf. Fig.(\ref{ocprob50})). 

To calculate the distribution of the $\lambda$ values for the trial state one can use the coefficients~(\ref{f1}). Fig.(\ref{lams}) shows the obtained distribution for $\gamma=-1.5$, $N=6$, and $\Delta=0.2$. Notice that there are approximately six different values of $\lambda$ with significative values in the probability distribution. The maximum is centered between $\lambda=1$ and $\lambda=2$; adjusting a gaussian distribution we get a mean $\mu=1.87$ and a standard deviation $\sigma=2.06$. For reference, the exact diagonalization gives $\lambda_{0}=2$ for this value of $\gamma$. While only values of $\lambda$ up to 10 were considered in this chart, these amount to 99.979\% of the total contribution. This is the main reason why the tensorial product of coherent states constitutes a good approximation to the quantum ground state.

\begin{figure}[h]
\scalebox{1.0}{\includegraphics{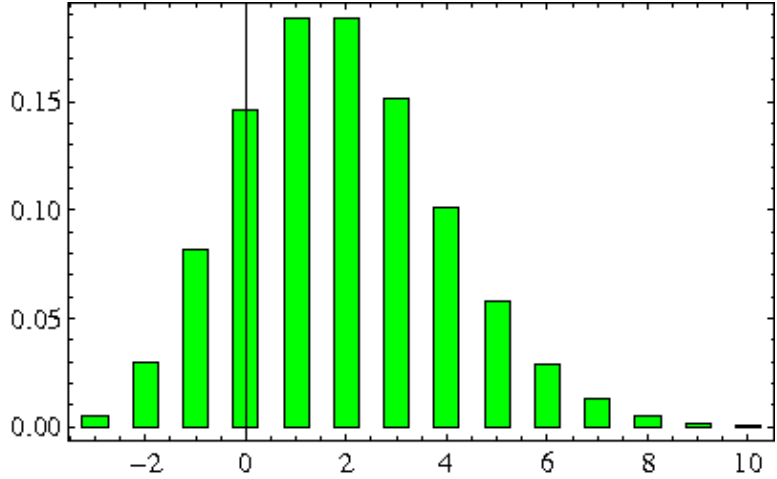}}
\caption{\label{lams} $\lambda$ distribution of the variational state, for $N=6$ atoms, $\Delta=0.2$, and $\gamma=-1.5$. It is approximately a gaussian distribution with mean $\mu=1.87$ and standard deviation $\sigma=2.06$. The exact diagonalization gives $\lambda_{0}=2$ for this value of $\gamma$.}
\end{figure}

\begin{figure}[h]
\scalebox{0.55}{\includegraphics{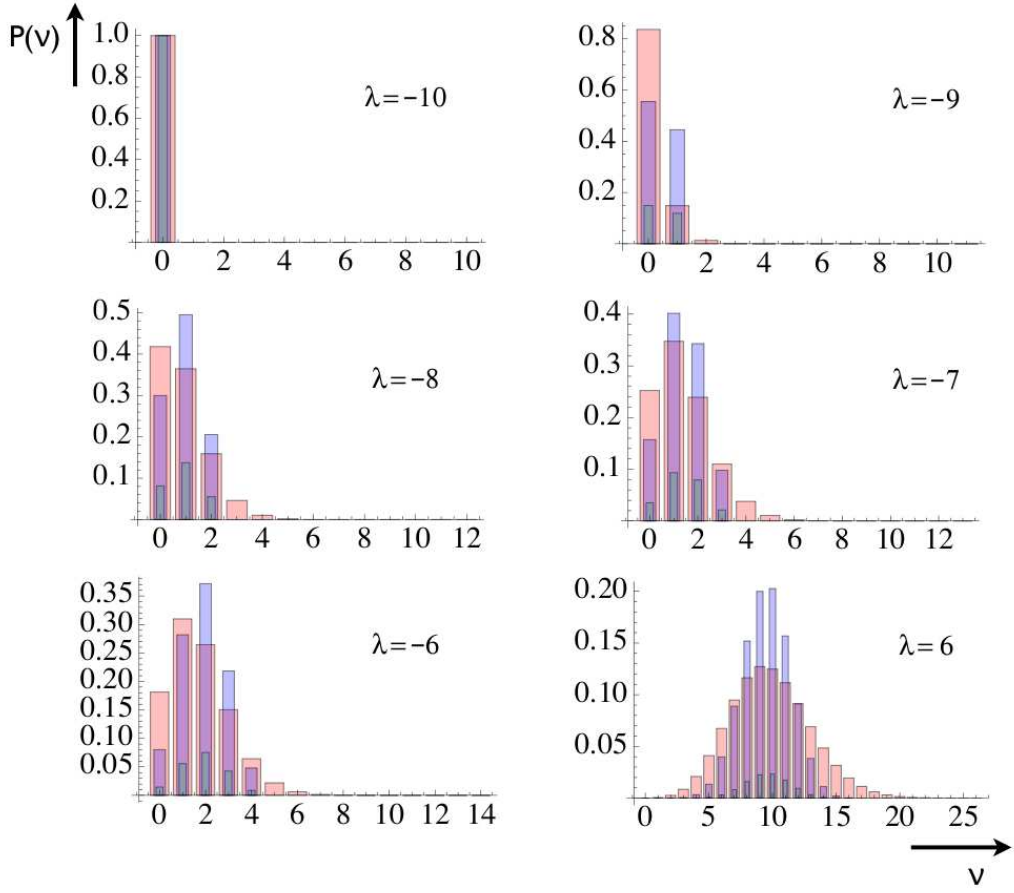}}
\caption{\label{traslapes} ({\it color online}) Photon number distribution for different eigenvalues of the constant of motion $\lambda$, for the exact quantum ground state with $N=20$ particles and $\Delta=0.2$, represented by the medium-width (blue) bars. Starting at the minimum value $\lambda=-10$, corresponding to $\gamma=\sqrt{1-\Delta }-0.01$ (just inside the North Pole), we increase $\lambda$ in steps of $1$, crossing the separatrix and reaching a value away from it at $\lambda=6$, corresponding to $\gamma=1.5$. In wider bars (magenta) the corresponding distribution for the variational case, containing all values of $\lambda$, is shown. If we consider the trial state with the contribution of only the $\lambda$ value given by the exact solution, the narrow (darker, green) bars result.}
\end{figure}

\begin{figure}[h]
\scalebox{0.55}{\includegraphics{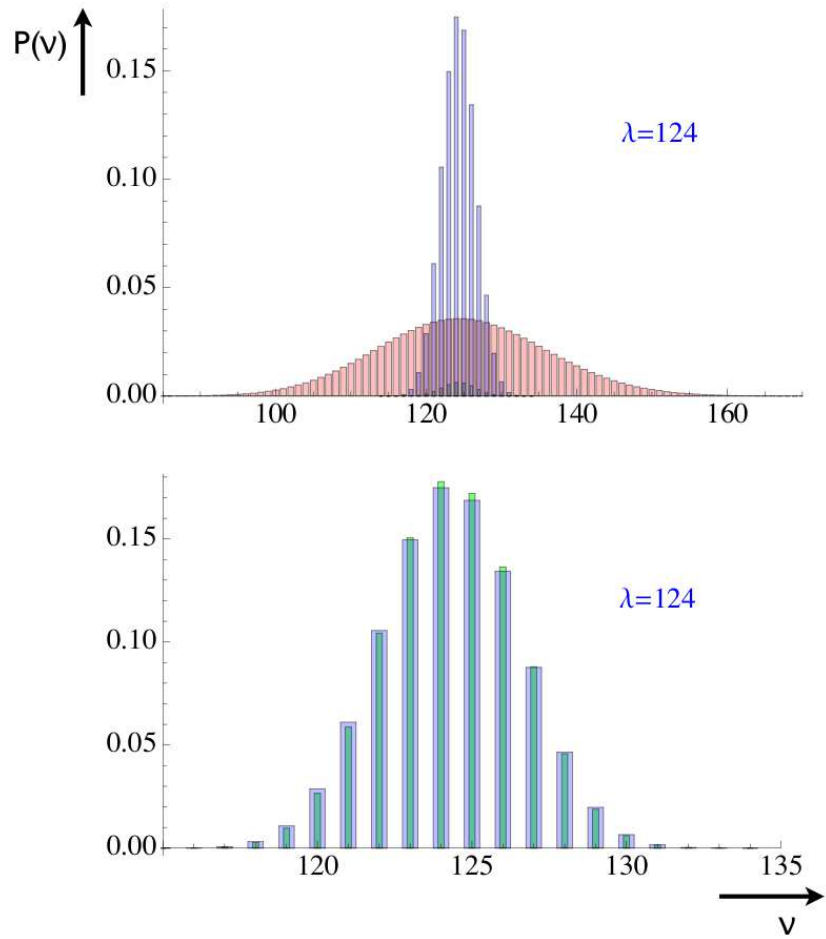}}
\caption{\label{traslapes2} Above: Same as Fig.(\ref{traslapes}) for $\lambda=124$, corresponding to $\gamma=5$, very far away from the separatrix. Below: Comparison of the renormalized trial state distribution (narrow bars) with the quantum distribution, for the same values.}
\end{figure}

In order to explain the good approximation of the trial state in all the studied observables except for the fluctuations in the number of photons, one can study the way in which the number of photons is distributed in the ground state for each quantum eigenvalue $\lambda$. From our analysis on fidelity (cf. Fig.(\ref{fidelity})), we expect the distributions for the variational state and the quantum state to agree inside the North Pole and away from the separatrix, but to show a strong disagreement at the separatrix itself. This is exactly what Fig.(\ref{traslapes}) shows, where, starting at the minimum value $\lambda=-10$, corresponding to $\gamma=\sqrt{1-\Delta }-0.01$ (just inside the North Pole), we increase $\lambda$ in steps of $1$, crossing the separatrix and reaching a value away from it at $\lambda=6$, corresponding to $\gamma=1.5$. We see that, even though the photon number distribution of the trial state is wider than the quantum case, the mean is the same. The top of Fig.(\ref{traslapes2}), for $N=10,\ \Delta=0.2$ and $\gamma=5$ (corresponding to $\lambda=124$), shows this more clearly. Considering the contribution to the photon number probability distribution given by only the $\lambda$ value of the exact quantum solution, we obtain the narrow, darker bars in Figs.(\ref{traslapes}, \ref{traslapes2}). The amplitudes are necessary smaller since we have eliminated all contributions from other $\lambda$ values: for instance, with $N=10,\ \Delta=0.2$ and $\gamma=5$ the contribution of the values of $\nu$ to the state is of only $3.5\%$. However, a proper renormalization renders this distribution practically equal to that of the corresponding quantum photon number distribution, as illustrated in Fig.(\ref{traslapes2}) (bottom).

\clearpage

\section{Conclusions}

A comprehensive study of the ground state quantum phase transitions in the Tavis-Cummings model, for any number of atoms, has been presented as a function of the strength of the interaction $\gamma$ and the atom hyperfine two-level separation energy $\omega_A$. The minima of the expectation value of the Hamiltonian were calculated for a coherent trial state, the separatrix that divides the control parameter space was found, and the structure of the ground state of the system in all these regions was studied. 

First and second order phase transitions were determined as the separatrix is crossed, and their influence on the behaviour of observables of interest for the matter and the field were explored. These include the population and dipole moments of the two level atoms, the expectation value of the number of photons and its fluctuations, and the squeezing and entanglement properties of the ground state. Additional quantum phase transitions of the ground state of the model, due to the discreteness of the value of the constant of motion $\lambda_0$, disappear when the number of atoms grows to infinity. However, the expectation values of the field and matter observables, together with the entanglement entropy, are manifestly discrete even for a very large number of atoms.

A discrete behaviour in the expectation value of the constant of motion $\Lambda$ at which the minimum energy is obtained, reflecting the quantum nature of the system, was shown: as the matter-field coupling is increased, the value of $\lambda_{0}$ remains constant until a threshold is reached, at which point it jumps to its next value. The discrete behaviour is inherited in all other quantum observables of interest. Since the value of observables remain constant over small regions of the interaction parameter, this feature can be used for control and manipulation of quantum systems of small number of atoms, where non-demolition experiments could be carried out. It was also found that when the detuning parameter is taken away from resonance, the $\lambda_{0}$ steps (and those of other observables) acquire a slope given precisely by the difference in energy between the absorbed photons and the atom's energy level separation. The effect of the detuning is of an even greater importance, as it modifies expectation values of interest such as the number of photons, the entanglement entropy, and the dipole moments. Our calculations show that these increase as $\Delta$ decreases, and viceversa. Thus, one could fine-tune the entanglement entropy $S_E$ between atoms, for instance, through a fine-tuning of $\Delta$.

While their semiclassical counterparts show, in contrast, a continuous behaviour, we have shown that they are an excellent approximation to the exact (quantum) values in all the calculated observables, with the exception of the photon number fluctuation. The reason for this is found in the fact that the proposed trial function, as a coherent state for the description of the electromagnetic field, includes an infinite number of photons in its composition, greatly overestimating the quantum result. This results in the trial state having a very small overlap with the exact quantum state, which may also be seen in the squeezing coefficient, as it exceeds the value of one for large values of the coupling parameter (cf. Fig.(\ref{sqe})). In reality, the variational state has a gaussian distribution of $\lambda's$ around the quantum value of the constant of motion. In the case of $N=6$ the standard deviation of this distribution is $\sigma=2$, implying that significant contributions will be had only for a small number of occupation states, or equivalently, from a small number of Fock states. A better trial state may be constructed by truncating the proposed state to only one value of $\lambda$, which can be chosen to be in agreement with that of the quantum solution, and by renormalizing the state. When this is done, we have found an excellent agreement between this renormalized trial state and the quantum result (cf. Fig.(\ref{traslapes2})). The calculation of the expected values of other observables, using the renormalized state, is undergoing.

The composition of the ground state was shown in terms of the distribution of the atoms into the two hyperfine levels, both for the exact energy ground state and for its variational approximation. We showed that away from the separatrix the semiclassical occupation probability estimates very well the quantum result. The estimate is even better as the number of atoms $N$ increases, until they become indistinguishable. However, as we get close to the separatrix the semiclassical approximation gets poorer. The fidelity of the reduced probability distributions of the atoms populating the excited hyperfine level is calculated for the semiclassical and exact ground states, which gives information about the overlap in the region where a pure state dominates, i.e., $|\omega_{A}|\leq\gamma^{2}$. Outside this region we obtain information about the similarity between two mixed density matrices. This fidelity can also be fine-tuned through $\Delta$.

\acknowledgments

This work was partially supported by Conacyt-Mexico and DGAPA-UNAM. J.G.H. wishes to thank J. Dukelsky for interesting discussions.

\end{document}